\documentclass[aps,pra,10pt,letterpaper,twocolumn,tightenlines,superscriptaddress,notitlepage, nofootinbib, longbibliography]{revtex4-2}
\usepackage{qcircuit}
\usepackage[dvips]{graphicx}
\usepackage{amsmath,amssymb,amsthm,mathrsfs,amsfonts,dsfont}
\usepackage{subfigure, epsfig}
\usepackage{braket}
\usepackage{bm}
\usepackage{enumerate}
\usepackage{xcolor}
\usepackage{comment}
\usepackage{scalefnt}
\usepackage[
colorlinks = true,
citecolor = blue,
linkcolor = blue,
urlcolor = blue]{hyperref}

\usepackage{pagecolor}
\usepackage[T1]{fontenc}
\usepackage{lmodern}
\usepackage[utf8]{inputenc}
\usepackage{microtype}
\usepackage[capitalise]{cleveref}
\crefname{section}{Sec.}{Secs.}
\usepackage{tikzscale}
\usepackage{pgfplots}
\pgfplotsset{compat=newest}
\usetikzlibrary{plotmarks}
\usetikzlibrary{arrows.meta}
\usetikzlibrary{external}
\usepgfplotslibrary{patchplots}
\usepackage{grffile}

\usepackage{algorithm}
\usepackage{algpseudocode}

\newcommand{\tV}{\vert\kern-0.25ex\vert\kern-0.25ex\vert}

\DeclareMathOperator{\tr}{Tr}
\DeclareMathOperator{\Tr}{Tr}

\definecolor{applegreen}{rgb}{0.55, 0.71, 0.0}

\newcommand{\comments}[1]{}

\graphicspath{{./figure/}}

\newcommand{\coleq}{\mathrel{\mathop:}\nobreak\mkern-1.2mu=}

\newcommand{\mc}{\mathcal}

\newcommand{\mr}{\mathrm}

\newcommand{\fid}{\mathrm{fid}}
\newcommand{\ketbra}[2]{\vert #1 \rangle \! \langle #2 \vert}

\definecolor{almond}{rgb}{0.98, 0.81, 0.69}

\newtheorem{theorem}{Theorem}

\newtheorem{lemma}{Lemma}

\definecolor{mycolor1}{rgb}{0.98431,0.41569,0.29020}%
\definecolor{mycolor2}{rgb}{0.45490,0.66275,0.81176}%
\definecolor{mycolor3}{rgb}{0.87059,0.17647,0.14902}%
\definecolor{mycolor4}{rgb}{0.16863,0.54902,0.74510}%
\definecolor{mycolor5}{rgb}{0.64706,0.05882,0.08235}%
\definecolor{mycolor6}{rgb}{0.01569,0.35294,0.55294}%

\definecolor{divergent1}{rgb}{0.1680, 0.5117, 0.7266}%
\definecolor{divergent2}{rgb}{0.6680, 0.8633, 0.6406}%
\definecolor{divergent3}{rgb}{0.9883, 0.6797, 0.3789}%
\definecolor{divergent4}{rgb}{0.8398, 0.0977, 0.1094}%

\begin{document}

\title{Neural--Shadow Quantum State Tomography}

\author{Victor Wei}
\email{victor.wei203@gmail.com}
\affiliation{Department of Physics, McGill University, Montreal, QC, Canada}
\affiliation{Institute for Quantum Computing, University of Waterloo, Waterloo, ON, Canada}
\author{W. A. Coish}
\email{william.coish@mcgill.ca}
\affiliation{Department of Physics, McGill University, Montreal, QC, Canada}
\author{Pooya Ronagh}
\email{pooya.ronagh@uwaterloo.ca}
\affiliation{Institute for Quantum Computing, University of Waterloo, Waterloo, ON, Canada}
\affiliation{Department of Physics \& Astronomy, University of Waterloo, Waterloo, ON, Canada}
\affiliation{Perimeter Institute for Theoretical Physics, Waterloo, ON, Canada}
\affiliation{1QB Information Technologies (1QBit), Vancouver, BC, Canada}
\author{Christine A. Muschik}
\email{christine.muschik@uwaterloo.ca}
\affiliation{Institute for Quantum Computing, University of Waterloo, Waterloo, ON, Canada}
\affiliation{Department of Physics \& Astronomy, University of Waterloo, Waterloo, ON, Canada}
\affiliation{Perimeter Institute for Theoretical Physics, Waterloo, ON, Canada}

\begin{abstract}
Quantum state tomography (QST) is the art of reconstructing an unknown quantum state through measurements. It is a key primitive for developing quantum technologies. Neural network quantum state tomography (NNQST), which aims to reconstruct the quantum state via a neural network ansatz, is often implemented via a basis-dependent cross-entropy loss function. State-of-the-art implementations of NNQST are often restricted to characterizing a particular subclass of states, to avoid an exponential growth in the number of required measurement settings. To provide a more broadly applicable method for efficient state reconstruction, we present ``neural--shadow quantum state tomography'' (NSQST)---an alternative neural network-based QST protocol that uses infidelity as the loss function. The infidelity is estimated using the classical shadows of the target state. Infidelity is a natural choice for training loss, benefiting from the proven measurement sample efficiency of the classical shadow formalism. Furthermore, NSQST is robust against various types of noise without any error mitigation. We numerically demonstrate the advantage of NSQST over NNQST at learning the relative phases of three target quantum states of practical interest, as well as the advantage over direct shadow estimation. NSQST greatly extends the practical reach of NNQST and provides a novel route to effective quantum state tomography.
\end{abstract}
\date{\today}
\maketitle

\section{Introduction} \label{sec:Intro}

Efficient methods for state reconstruction are essential in the development of advanced quantum technologies. Important applications include the efficient characterization, readout, processing, and verification of quantum systems in a variety of areas ranging from quantum computing and quantum simulation to quantum sensors and quantum networks \cite{eisert2020quantum,knill2008randomized, carrasco2021theoretical, choi2023preparing, elben2023randomized, stricker2022experimental}. However, with physical quantum platforms growing larger in recent years \cite{preskill2018quantum}, reconstructing the target quantum state through brute-force quantum state tomography (QST) has become much more computationally demanding due to an exponentially increasing number of required measurements. To address this issue, various approaches have been proposed that are efficient in both the number of required measurement samples and in the number of parameters used to characterize the quantum state. These include classical shadows \cite{huang2020predicting} and neural network quantum state tomography (NNQST) \cite{torlai2018neural}. The goal of NNQST is to produce a neural network representation of a complete physical quantum state that is close to some target state. In contrast, the classical shadows formalism does not aim to reconstruct a full quantum state, but rather to obtain a reduced classical description that allows for efficient evaluation of certain observables.

A neural network quantum state ansatz has been shown to have sufficient expressivity to represent a wide range of quantum states \cite{huang2021neural,sharir2022neural,chen2018equivalence,deng2017quantum} using a number of model parameters that scales polynomially in the number of qubits. Furthermore, as methods for training neural networks have long been investigated in the machine learning community, many useful strategies for neural network model design and optimization have been directly adopted for NNQST \cite{hinton2012practical,vaswani2017attention,KingBa15}. Following the introduction of neural network quantum states \cite{carleo2017solving}, Torlai \emph{et al.}~proposed the first version of NNQST, an efficient QST protocol based on a restricted Boltzmann machine (RBM) neural network ansatz and a cross-entropy loss function \cite{torlai2018neural}. NNQST has been applied successfully to characterize various pure states, including W states, the ground states of many-body Hamiltonians, and time-evolved many-body states \cite{torlai2018neural,bennewitz2022neural,torlai2019integrating}. Despite the promising results of NNQST in many use cases, the protocol faces a fundamental challenge: An exponentially large number of measurement settings is required to identify a general unknown quantum state (although a polynomial number is sufficient in some examples \cite{chen2013uniqueness}). During NNQST, a series of measurements is performed in random local Pauli bases $B$ for $n$ qubits $(B=( P_1, P_2, \cdots, P_n )$, where $P_i\in\left\{X,Y,Z\right\}$). Because this set is exponentially large, some convenient subset of all possible $B$ must be selected for a large system, but this subset may limit the ability of NNQST to identify certain states. An important example is the phase-shifted multi-qubit GHZ state, relevant to applications such as quantum sensing. In this case, the relative phase associated with non-local correlations cannot be captured by measurement samples from almost-diagonal local Pauli bases, i.e., bases with $m\ll n$ indices $i$ for which $P_i = X$ or $P_i = Y$. Nonetheless, this limited set of almost-diagonal Pauli bases is widely used in NNQST implementations to avoid an exponential cost in classical post-processing \cite{torlai2018neural,bennewitz2022neural}.

To address this challenge, we use the classical shadows of the target quantum state to estimate the infidelity between the model and target states. This is in contrast with approaches that use the conventional basis-dependent cross-entropy as the training loss for the neural network. This choice is motivated by two main factors. Firstly, infidelity is a natural candidate for a loss function compared to cross-entropy; the magnitude of the basis-dependent cross-entropy loss is in general not indicative of the distance between the neural network quantum state and the target state. Additionally, infidelity is the squared Bures distance \cite{luo2004informational}, a measure of the statistical distance between quantum states that enjoys metric properties such as symmetry and the triangle inequality. The infidelity is therefore a better behaved objective function for optimization. Secondly, the classical shadow formalism of Huang \emph{et al.} was originally developed to address precisely the scaling issues of brute-force QST \cite{huang2020predicting}. Instead of reconstructing the unknown state, shadow-based protocols, first proposed by Aaronson \cite{aaronson2018shadow}, predict certain properties of the quantum state with a polynomial number of measurements. Therefore, classical shadows provide the following two main advantages in our work: (i) they are provably efficient in the number of required measurement samples for predicting various observables (e.g. the infidelity), and (ii) there is no choice of measurement bases required and therefore no previous knowledge of the target state is assumed.

\begin{figure}[!htb]
\centering
\includegraphics[width = \columnwidth]{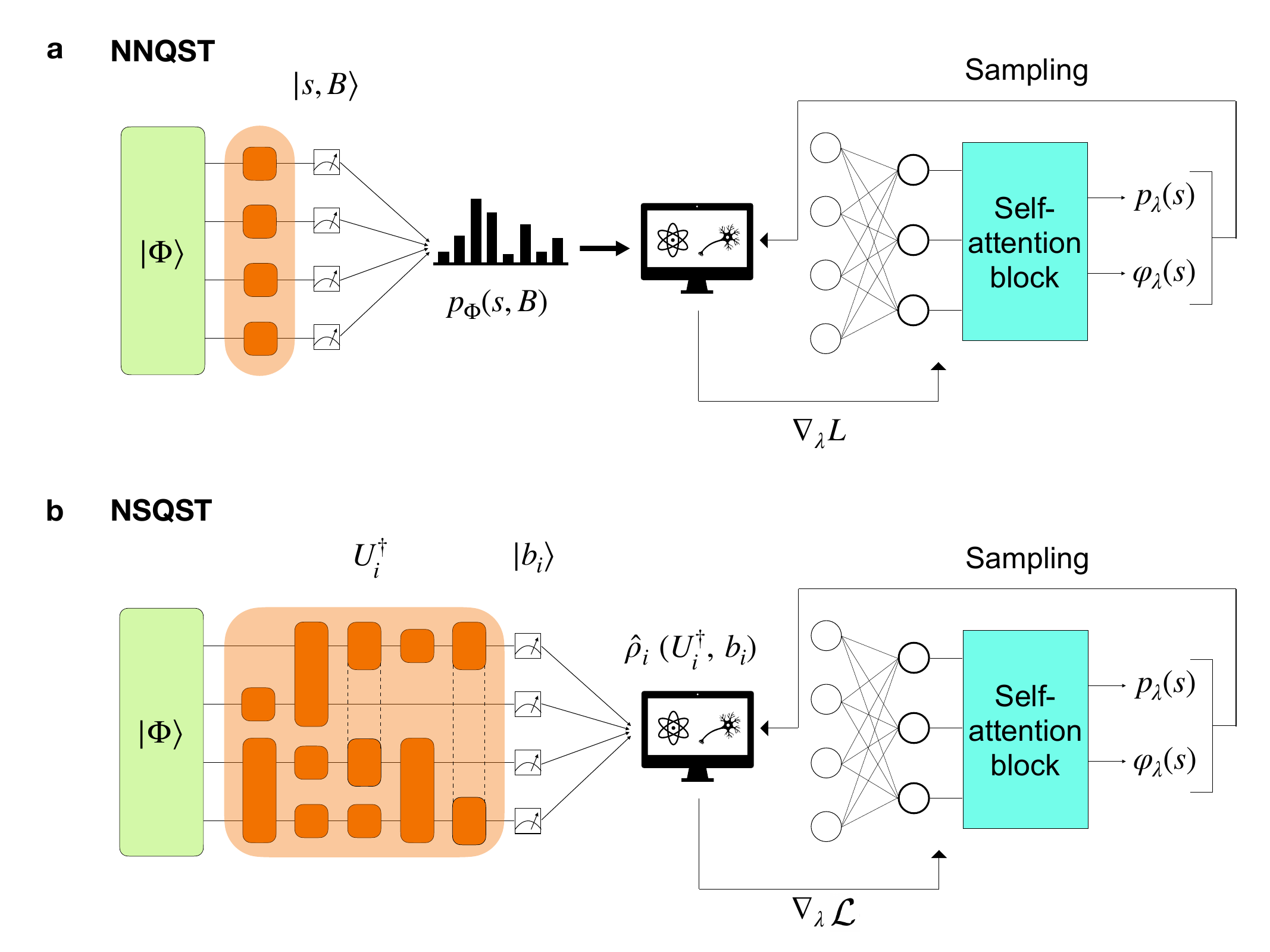}
\caption{Overview of the NNQST and NSQST protocols. Panel \textbf{a} shows the NNQST protocol with the cross-entropy loss function $L_{\lambda}$. The training data determine $p_{\Phi}(s,B)$, the measured probability distribution of measurement outcomes $s$ for measurements of the target state $\ket{\Phi}$ performed in the local Pauli basis $B$. The feedback loop on the right-hand side indicates the iterative first-order optimization for neural network training. Panel \textbf{b} displays the NSQST protocol described in \cref{standard_nsqst}, where the training data set consists of classical shadows only and where the network parameters $\lambda$ are trained via an infidelity loss function $\mathcal L_\lambda$. The expression $\hat{\rho}_{i} \ (U^{\dagger}_{i}, \ b_{i})$ is the stored classical shadow of the target state $\lvert\Phi\rangle$ with the Clifford unitary $U_i^{\dagger}$ and bit-string $\lvert b_{i}\rangle$.}
\label{fig:nsqst_overview}
\end{figure}

Our new pure-state QST protocol, ``neural--shadow quantum state tomography'' (NSQST), reconstructs the unknown quantum state in a neural network quantum state ansatz by using classical shadow estimations of the gradients of infidelity for training (\cref{fig:nsqst_overview}b). In our numerical experiments, NSQST demonstrates clear advantages in three example tasks: (i) reconstructing a time-evolved state in one-dimensional quantum chromodynamics, (ii) reconstructing a time-evolved state for an antiferromagnetic Heisenberg model, and (iii) reconstructing a phase-shifted multi-qubit GHZ state. Moreover, the natural appearance and inversion of a depolarizing channel from randomized measurements in the classical shadow formalism makes NSQST noise-robust without any calibration or modifications to the loss function, while one of these two extra steps is required in noise-robust classical shadows \cite{koh2022classical,chen2021robust}. We numerically demonstrate NSQST's robustness against two of the most dominant sources of noise across a wide range of physical implementations: two-qubit CNOT errors and readout errors. The rest of this paper is organized as follows: In \cref{sec:Methods}, we summarize the methods used in our numerical simulations, including the neural network quantum state ansatz, NNQST, classical shadows, NSQST, and NSQST with pre-training. In \cref{sec:noiseless_sim} and \cref{sec:noisy_sim}, we provide numerical simulation results for NNQST, NSQST, and NSQST with pre-training in three useful examples, both noise-free and in the presence of noise. In particular, we provide a comparison to direct shadow estimation in \cref{direct_shadow_sec}. Finally, \cref{sec:conclusions} summarizes the key advantages of NSQST and some possible future directions. We provide additional technical details and suggestions for further improvements to NSQST in the appendices.

\section{Methods} \label{sec:Methods}

In this section, we describe existing methods for characterizing quantum states and then introduce and describe two variants of NSQST. We introduce neural network quantum states in \cref{method_NN_state}. State-of-the-art NNQST implementations and the classical shadow protocol are summarized in \cref{method_NNQST} and \cref{method_shadow}, respectively. Our proposed NSQST protocol is described in \cref{standard_nsqst}. In addition, a modified NSQST protocol with pre-training is described in \cref{pre_nsqst}.

\subsection{Neural network quantum state}\label{method_NN_state}

Our pure-state neural network quantum state ansatz is adopted from Ref.~\cite{bennewitz2022neural}. The parameterized model is based on the transformer architecture \cite{vaswani2017attention}, widely used in natural language processing and computer vision \cite{wolf2020transformers,khan2022transformers}. As compared to older architectures such as the RBM, the transformer is superior in modelling long-range interactions and allows for more efficient sampling of the encoded probability distribution due to its autoregressive property \cite{bennewitz2022neural}.

The transformer neural network quantum state ansatz takes a bit-string $s = (s_1,\ldots,s_n) \in \{0,1\}^n$ corresponding to the computational basis state $|s\rangle$, and produces a complex-valued amplitude $\braket{ s | \psi_{ \lambda} }=\psi_{ \lambda} (s)$ parameterized by ${\lambda} = ({\lambda_{1}}, {\lambda_{2}})$ as
\begin{equation} \label{NNQST:nn_wave_fun}
 \psi_{ \lambda} (s)
  = \sqrt{p_{ \lambda_{1}}(s)} e^{i \varphi_{ \lambda_{2}}(s)},
\end{equation}
where $\lambda_{1}$ and $\lambda_{2}$ are vectors of real-valued model parameters for the normalized probability amplitudes $p_{ \lambda_{1}}(s)$ and the phases $\varphi_{ \lambda_{2}}(s)$ of the neural network quantum state. These amplitudes and phases may be parameterized by the neural network quantum state in various fashions. One approach is to use two completely disjoint models, independently parameterized by $\lambda_{1}$ and $\lambda_{2}$ for the amplitude and phase values, respectively \cite{torlai2018neural,torlai2018latent}. Another approach is to use a single model parameterized by $\lambda$ to encode both the amplitude and phase outputs, either via complex-valued model parameters \cite{carleo2017solving,vicentini2022netket} or by using real-valued model parameters with two disjoint layers of output neurons connected to a common preceding neural network \cite{bennewitz2022neural}. In our numerical experiments, we use the later parameterization for NNQST and NSQST, but the modified NSQST protocol with pre-training in \cref{pre_nsqst} uses two separately parameterized neural networks. See \cref{Appendix_nn} for a more detailed account of the transformer architecture.

Given a trained neural network quantum state, observables and other state properties of interest can be predicted by drawing (classical) samples from the neural network model. The number of samples required to predict the expectation value of an arbitrary Pauli string (independent of its weight) and fidelity to a computationally tractable state with bounded additive error is independent of the system size \cite{havlicek2023amplitude}. Computationally tractable states include stabilizer states and neural network quantum states. Thus, if sampling can be performed efficiently, the prediction errors from neural network quantum states are primarily due to imperfect training. We also note that not every neural network quantum state ansatz has this property of efficient observable and fidelity prediction, where an important example is a class of generative models trained on informationally complete positive-operator valued measures (IC-POVMs) \cite{carrasquilla2019reconstructing, cha2021attention}.

\subsection{Neural network quantum state tomography (NNQST)}\label{method_NNQST}

NNQST (\cref{fig:nsqst_overview}a) aims at obtaining a trained neural network representation that closely approximates an unknown target quantum state. The training is done by iteratively adjusting the neural network parameters along a loss gradient estimated from the measurement samples in various local Pauli bases (obtained by applying single-qubit rotations before performing measurements in the computational basis) \cite{bennewitz2022neural}. We denote a local Pauli basis as $B = (P_1, P_2, \cdots, P_n)$, with $P_i \in \{X, Y, Z\}$. If a measurement sample $s \in \{0, 1\}^n$ is obtained after performing rotations to the Pauli basis $B$, we store the pair $(s,B)$ as a training sample, corresponding to a product state $\ket{s,B}$.

After choosing a subset $\mathcal{B}$ of Pauli bases for collecting measurement samples, we estimate a loss function that represents the distance between the target state $\ket{\Phi}$ and the neural network quantum state $\ket{\psi_{ \lambda}}$. The loss function in NNQST is based on the cross-entropy of the measurement outcome distributions for the target and neural network states in each basis $B$, which is then averaged over the set of bases $\mathcal{B}$. Ignoring a $\lambda$-independent contribution arising from the average entropy of the target-state measurement distribution, this procedure gives the cross-entropy loss function for NNQST \cite{bennewitz2022neural}:
\begin{align} \label{NNQST_cost}
  L_{ \lambda} =
  -\frac{1}{|\mathcal{B}|} \sum_{B \in \mathcal B} \sum_{s \in \{0, 1\}^n}
  p_{\Phi}(s, B) \ln p_{\psi_{ \lambda}}(s, B).
\end{align}
Here, $p_{\Phi}(s, B)$ is the probability of measuring the outcome $s$ from the target state $\ket{\Phi}$ after rotating to the Pauli basis $B$ and $p_{\psi_{ \lambda}}(s, B)$ is defined as
\begin{align} \label{likelihood}
  p_{\psi_{ \lambda}}(s, B)
  &= \Big| \sum_{\substack{ t \in \{0,1\}^n \\ \langle s, B| t \rangle \neq 0}}
  \braket{ s, B| t }\!\braket{ t | \psi_{ \lambda} } ~ \Big|^2,
\end{align}
where the overlap between the Pauli product state and the neural network quantum state requires a summation over the computational basis states $\ket{t}$ that satisfy $\braket{ s, B| t } \neq 0$. Note that the number of these states $\ket{t}$ is $2^K$, with $K$ being the number of positions $i$ where $P_i \neq Z$. This suggests that an efficient and exact calculation of $p_{\psi_{ \lambda}}(s, B)$ requires the projective measurements to be in almost-diagonal Pauli bases for a generic neural network quantum state $\ket{\psi_\lambda}$ \cite{bennewitz2022neural}.

Using the law of large numbers, the cross-entropy loss can be approximated via a finite training data set $\mathcal{D}_{T}$ as
\begin{align} \label{NNQST_approx_cost}
  L_{ \lambda}
  \approx -\frac{1}{|\mathcal{D}_T|}
  \sum_{\ket{s, B} \in \mathcal D_T} \ln p_{\psi_{ \lambda}}(s, B).
\end{align}
An approximation for the gradient $\nabla_\lambda L$ is then directly found from \cref{NNQST_approx_cost}. During training, the gradient is provided to an optimization algorithm such as stochastic gradient descent (SGD) or one of its variants (e.g., the Adam optimizer \cite{KingBa15}). In this paper, we exclusively use the Adam optimizer.

\subsection{Classical shadows}\label{method_shadow}

Shadow tomography relies on the ingenious observation that a polynomial number of measurement samples is sufficient to predict certain observables for quantum states of arbitrary size \cite{aaronson2018shadow}. The classical shadow protocol further exploits the efficiency of the stabilizer formalism, making this procedure ready for practical experiments \cite{huang2020predicting,struchalin2021experimental, becker2024classical, zhang2021experimental}. In this paper, we focus on estimating linear observables of the form $\Tr(O\rho)$ for a pure state $\rho= \ketbra{\Phi}{\Phi}$. An important example (for $O=\ketbra{\psi_\lambda}{\psi_\lambda}$) is the fidelity between the target state $\ket{\Phi}$ and a reference state $\ket{\psi_\lambda}$.
The first step in the protocol is to collect the so-called classical shadows of $\ket{\Phi}$. To obtain a single classical shadow sample, we apply a randomly-sampled Clifford unitary $U_{i}\in{\sf Cl}(2^n)$ to the quantum state and measure all $n$ qubits in the computational basis, resulting in a single bit-string $\ket{b_i}$. The stabilizer states $\ket{\phi_i}= U_i^\dagger \ket{b_i}$ contain valuable information about $\rho$. Using representation theory \cite{Webb2015TheCG}, it can be shown that the density matrix obtained from an average over both the random unitaries and the measured bit-strings $\mathcal M(\ketbra{\Phi}{\Phi}):= \mathbb E_{U \sim {\sf Cl}(2^n), b \sim P_{\Phi}(b)} [\ketbra{\phi}{\phi}]$ coincides with the outcome of a depolarizing noise channel:
\begin{equation}\label{depolarizing_channel}
\begin{aligned}
\mathcal M(\rho)= \mathcal D_{n,1/(2^n+1)} (\rho),
\end{aligned}
\end{equation}
where $\mathcal D_{n,f} (\rho)= f \rho + (1-f) \frac{\mathbb I}{2^n} $ denotes the depolarizing noise channel of strength $f$. The original state can then be recovered as an average over classical shadows by inverting the above formula, $\ketbra{\Phi}{\Phi}=\mathbb E\left[\mathcal M^{-1} (\ketbra{\phi}{\phi})\right]$. We emphasize that the original state can only be recovered after sampling from a prohibitively large number of Clifford unitaries, and for each of them sampling an exponentially large number of bit-strings. The classical shadows are therefore defined by \cite{huang2020predicting}:
\begin{equation}\label{def_shadow_noiseless}
\begin{aligned}
\hat{\rho}_{i}(U_i, b_i)
:= \mathcal M^{-1} (\ketbra{\phi_i}{\phi_i})
= (2^n + 1) \ketbra{\phi_i}{\phi_i} - \mathbb{I}.
\end{aligned}
\end{equation}

More generally, in the presence of a gate-independent, time-stationary, and Markovian noise channel $\mathcal E$ afflicting the segment of the circuit between the preparation of the state $\rho$ and the measurements, this definition extends to \cite{koh2022classical}:
\begin{equation}\label{def_shadow_noisy}
\begin{aligned}
\hat{\rho}_{i}(\mathcal E, U_i, b_i) = \frac{1}{f(\mathcal E)} \ketbra{\phi_i}{\phi_i} + \Big(1 - \frac{1}{f(\mathcal E)}\Big)\frac{\mathbb{I}}{2^n}.
\end{aligned}
\end{equation}
Here, $f(\mathcal E)$ is the strength of a depolarizing noise channel $\mathcal D_{n,f(\mathcal E)}$ comprised of the combined effects of the channel in \cref{depolarizing_channel} and the twirling of the additional noise by the random Clifford unitaries, effectively imposing further depolarization. Koh and Grewal \cite{koh2022classical} derived an analytic expression for $f(\mathcal E)$ as
\begin{align} \label{Shadow:f_eps}
f(\mathcal E) = \frac{\fid(\mathcal E)-1}{2^{2n}-1},
\end{align}
where
\begin{align} \label{Shadow:f_eps_def}
\fid(\mathcal E)= \Tr(\mathcal E\circ \mathrm{diag}) = \sum_{s \in \{0, 1\}^n } \bra s \mathcal E(\ketbra ss) \ket s
\end{align}
is the sum of fidelities for the noise channel $\mathcal E$ acting on each of the computational basis states $\ket{s}$ ($\fid(\mathcal E)\in [0,2^n]$; at the lower bound, $\fid(\mathcal E)=0$, the depolarizing parameter becomes negative, $f(\mathcal E)<0$, but the associated depolarizing channel remains physical \cite{koh2022classical}). When the noise channel is not exactly known, extra calibration procedures are required in noise-robust classical shadow protocols \cite{chen2021robust}.

Once the classical shadow samples $\{\hat{\rho}_i\}_{i=1}^N$ are collected, we calculate $\hat o^{(i)}\coleq \Tr(\hat{\rho}_{i}O)$ for each of the $N$ classical shadows and obtain an estimator for the observable from an average over the $N$ samples (alternatively, the median-of-means can be used to improve the success rate of the protocol; see Ref.~\cite{huang2020predicting} for more details). The key advantage of classical shadows is the bounded variance of observable estimations which, in turn, provides a bound on the number of classical shadow samples required to predict linear observables of the quantum state within a target precision. Indeed, as shown in Refs.~\cite{chen2021robust,koh2022classical}, the number of classical shadow samples $N$ required to estimate $M$ arbitrary linear observables $\{O_j\}_{j=1}^M$, up to an additive error $\varepsilon$, scales as
\begin{align} \label{Shadow:scaling_noise}
N &=\mc O\left (\frac{2^{2n}\log M}{\varepsilon^{2}\fid(\mathcal E)^{2}} \max_{1 \leq j \leq M} \tr(O_j^2) \right).
\end{align}

In the case of noise-free Clifford tails, $\mathcal E = \mathbb I$, we have $\fid(\mathbb I) = 2^n$, $f(\mathbb I)= (1+2^n)^{-1}$, and we recover the sample complexity
$\mc O\left (\max_{1 \leq j \leq M} \tr(O_j^2) \log M/\varepsilon^{2}  \right)$
presented in \cite{huang2020predicting}, which is indeed independent of the system size $n$. The variance of observable estimators is also bounded as $\mr{Var}(\hat o) \leq 3\Tr(O^2)$ in this case and is independent of the system size.

\subsection{Neural--shadow quantum state tomography (NSQST)}\label{standard_nsqst}

Given a pure target state $\ket{\Phi}$, our goal in NSQST (\cref{fig:nsqst_overview}b) is to progressively adjust the model parameters ${\lambda}$, such that the associated pure state $\ket{\psi_{ \lambda}}$ (see \cref{NNQST:nn_wave_fun}) approaches $\ket{\Phi}$ during optimization. We approximate the fidelity between the model and target states using the classical shadow formalism described in \cref{method_shadow}, taking $O_\lambda = \lvert \psi_{ \lambda}\rangle\!\langle \psi_{ \lambda}\rvert$ as a linear observable, averaged with respect to $\rho = \lvert \Phi\rangle\!\langle \Phi\rvert$. The number $M$ of observables we predict during optimization therefore coincides with the number of descent steps taken by the optimizer, as updating $\lambda$ changes the observable $\lvert \psi_{ \lambda}\rangle\!\langle \psi_{ \lambda}\rvert$ in every iteration. By collecting $N$ classical shadows, we can approximate our loss function (the infidelity) via
\begin{equation}\label{nsqst_loss_noisy}
\begin{aligned}
\mathcal L_\lambda(\mathcal E)&:= 1-|\langle \psi_{ \lambda} \rvert \Phi\rangle|^2 \\
&\approx 1- \frac{1}{N}\sum_{i=1}^{N}\Tr(O_\lambda\hat{\rho}_{i}) \\
&=1- \frac{1}{N}\sum_{i=1}^{N} \bra{\psi_{ \lambda}} \hat{\rho}_{i}(\mathcal E, U_i, b_i) \ket{\psi_{ \lambda}}\\
&=1- \frac{1}{2^n}\left(1 - \frac{1}{f(\mathcal E)}\right) - \frac{1}{Nf(\mathcal E)}\sum_{i=1}^{N} |\braket{\phi_i|\psi_{ \lambda}}|^2.
\end{aligned}
\end{equation}
In the noise-free case $\mathcal E = \mathbb I$, this expression simplifies to
\begin{equation}\label{nsqst_loss_noiseless}
\begin{aligned}
\mathcal L_\lambda(\mathbb I)\approx
2 - \frac{2^n+1}{N}\sum_{i=1}^{N} |\braket{\phi_i|\psi_{ \lambda}}|^2.
\end{aligned}
\end{equation}
We see that, independent of the specific form of $\mathcal E$, training the model is simply equivalent to increasing the average overlap between the random stabilizer states and the model quantum state.

The next step in NSQST requires classical post-processing to estimate the overlaps $\braket{\phi_i|\psi_{ \lambda}}$. For certain states $\ket{\psi_\lambda}$ (e.g., stabilizer states), the overlap can be calculated efficiently. Many states of interest do not fall into this class, leading to a potential exponential overhead. However, we can obtain a Monte-Carlo estimate of the overlap by sampling from the model quantum state $\ket{\psi_\lambda}$. In the model, we associate a probability
\begin{equation}\label{nqs_sample}
\begin{aligned}
p_\lambda(s)= |\psi_{ \lambda}(s)|^2
\end{aligned}
\end{equation}
to each computational basis state $\ket{s}$. Therefore,
\begin{equation}\label{inner_prod}
\begin{aligned}
&\langle \phi_{i}\ket{ \psi_{ \lambda}} = \sum_{s}\frac{\phi_{i}^{*}(s)}{\psi_{ \lambda}^{*}(s)} p_\lambda(s) = \Big\langle \frac{\phi_{i}^{*}(s)}{\psi_{ \lambda}^{*}(s)} \Big\rangle_{\psi_\lambda}.
\end{aligned}
\end{equation}
It is now straightforward to provide a Monte Carlo estimate of the above quantity \cite{choo2018symmetries} (see \cref{app_technical} for an alternative approach). For each sample $s$ from the neural network, we have direct access to the exact complex-valued amplitudes $\psi_\lambda(s)$ of the neural network quantum state in the computational basis. Moreover, we can compute the stabilizer state projections $\phi_i(s)= \bra{s} U_i^\dagger \ket{b_i}$ in $O(n^2)$ time, in view of the Gottesman-Knill theorem \cite{gottesman1997stabilizer, gidney2021stim, aaronson2004improved}. Note that decomposing a randomly sampled Clifford operator into primitive unitary gates (e.g., with Hadamard, S, and CNOT gates) still takes $O(n^3)$ time \cite{aaronson2004improved, bravyi2021hadamard}. However, this is a one-time procedure to be run for each $U_i$ and can be done in advance of state tomography.

For first-order optimization methods (such as SGD and Adam), it is the gradient of the loss function rather than the loss function itself that must be estimated. From \cref{nsqst_loss_noisy} and using the log-derivative trick, we obtain the gradient
\begin{equation}\label{nsqst_grad}
\begin{aligned}
\nabla_\lambda \mathcal L(\mathcal E) \approx \frac{-2}{N f(\mathcal E)} \sum_{i=1}^{N} \Re\Big[ \Big\langle\frac{\phi_{i}^{*}(s)}{\psi_{\lambda}^{*}(s)}D_{\lambda}(s) \Big\rangle_{\psi_\lambda}\Big\langle\frac{\phi_{i}(s)}{\psi_{\lambda}(s)} \Big\rangle_{\psi_\lambda} \Big],
\end{aligned}
\end{equation}
where we define the diagonal operator $D_\lambda$ as
\begin{equation}\label{grad_O_op}
\begin{aligned}
D_\lambda(s) &= \frac{1}{\psi_{ \lambda}(s)}\frac{\partial \psi_{ \lambda}(s)}{\partial \lambda} = \nabla_{ \lambda} \ln{\langle{s}\ket{\psi_{ \lambda}}} \\ &= \nabla_{ \lambda} \Big(\ln{\sqrt{p_{ \lambda}(s)}} + i\varphi_{ \lambda}(s)\Big).
\end{aligned}
\end{equation}
A simple but important observation is that the noise enters \cref{nsqst_grad} only in the overall prefactor $\propto 1/f(\mathcal E)$. Thus, the noise may affect the learning rate, but it will not affect the direction of the gradient. This suggests that gradient-based optimization schemes can yield an accurate neural network quantum state without any noise calibration or mitigation. This is despite the fact that this same noise generally biases the estimated infidelity (see \cref{nsqst_loss_noisy}).

We now discuss a possible limitation of our approach to classical post-processing. Given $N$ classical shadows collected experimentally, and $L$ computational basis samples collected from the neural network quantum state, the number of Monte Carlo samples collected from the neural network quantum state must be $L\sim O\left(1/N^2 f(\mathcal E)^2\right)$ in order to guarantee a bounded standard error in the approximation of the gradient from \cref{nsqst_grad}. Since $f(\mathcal E)\le 1/(1+2^n)$, this suggests that there may be an exponential cost in performing the Monte Carlo estimations. We emphasize that this potential exponential cost in classical post-processing does not affect the required number of classical shadows from measurements. As system sizes grow significantly larger, it will eventually become hopeless to perform an exact sum over all $2^n$ computational basis states and the Monte Carlo average may still lead to successful convergence with only a sub-exponential number of samples in some cases (further details and an alternative approach to performing the Monte Carlo average are discussed in \cref{app_technical}). In our numerical simulations with six qubits (see \cref{sec:noiseless_sim}), having $2^6=64$ computational basis states, we have evaluated \cref{nsqst_grad} using 5000 Monte Carlo samples. With this many samples, the Monte Carlo estimation error is negligible and statistical fluctuations in the gradient are predominantly due to the finite number of classical shadows collected. 

\subsection{NSQST with pre-training}\label{pre_nsqst}
\begin{figure}[!htb]
\centering
\includegraphics[width = \columnwidth]{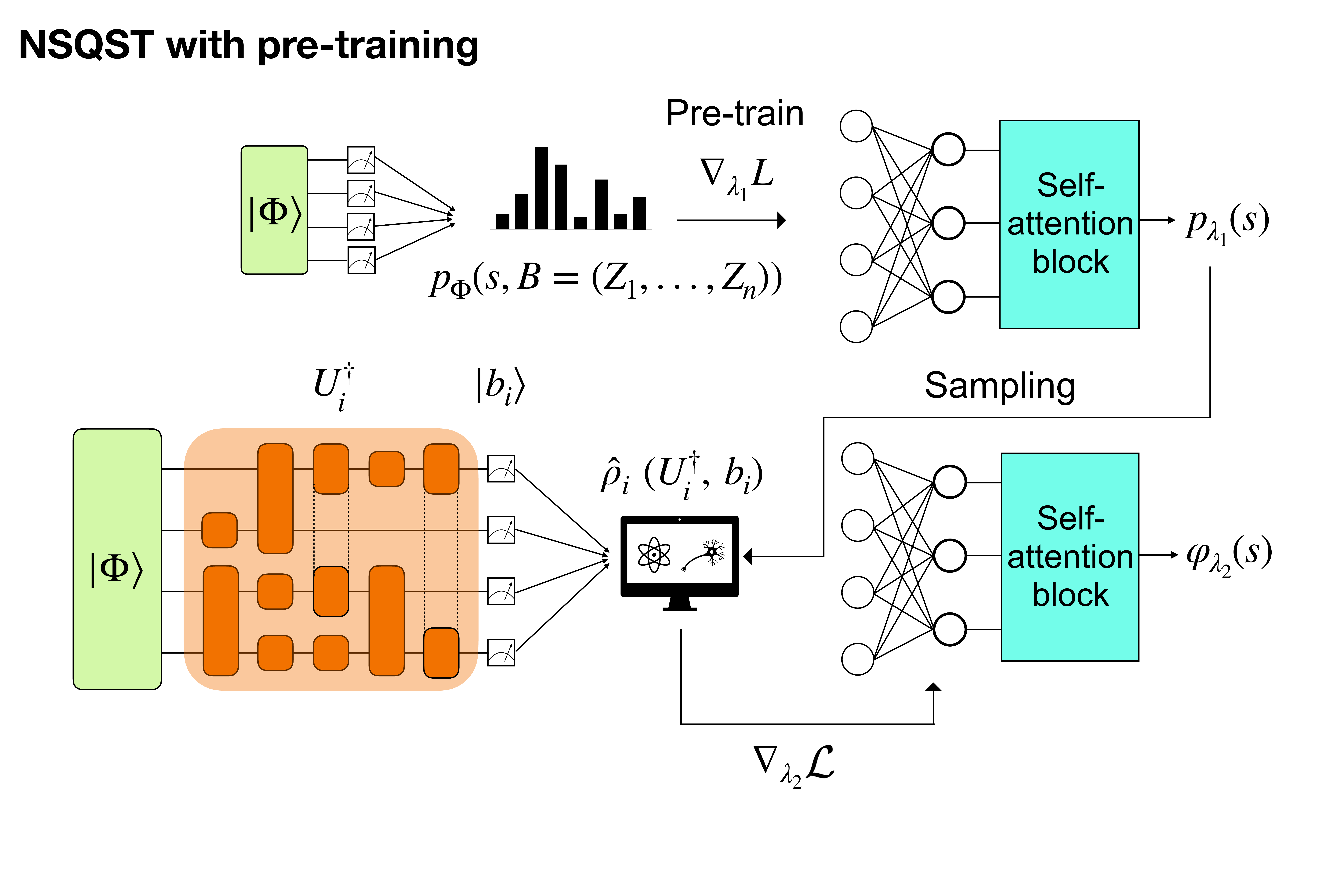}
\caption{Overview of the NSQST with pre-training protocol. The neural networks learning $p_{ \lambda_{1}}(s)$ and $\phi_{ \lambda_{2}}(s)$ are separately parameterized, and $p_{ \lambda_{1}}(s)$ is pre-trained using NNQST with training data derived from measurements only in the computational basis.}
\label{fig:pre_nsqst_overview}
\end{figure}
Along with the standard NSQST protocol, we also outline a modified NSQST protocol we call ``NSQST with pre-training'', which combines the resources used in NNQST and the standard NSQST protocol. NSQST with pre-training aims to find a solution with a lower infidelity than either of the other protocols alone. In this protocol we train two models with disjoint sets of parameters $\lambda_1$ and $\lambda_2$. We call these models the \emph{probability amplitude model} and the \emph{phase model}. \Cref{fig:pre_nsqst_overview} provides a visual overview of the protocol for NSQST with pre-training.

First, the parameters $\lambda_1$ are optimized to produce an accurate distribution $p_{ \lambda_1}(s)\simeq p_\Phi(s,B)$ from measurements performed exclusively in the computational basis [$B = (Z_1, Z_2,\cdots, Z_n)$]. Note that we can efficiently evaluate the loss function, \cref{NNQST_approx_cost}, and its gradient in this case as they depend only on the probabilities and not on the phases. Next, we perform NSQST to train the model parameters $\lambda_2$, learning the phases $\varphi_{\lambda_2}(s)$. However, unlike the case of standard NSQST, to perform a  Monte Carlo estimate of the gradient, \cref{nsqst_grad}, here we select random samples from a set of computational basis states $s$ according to the pre-trained distribution $p_{\lambda_1}(s)$. Since the NSQST Monte Carlo approximations do not follow the model $\lambda_2$ in an \emph{on policy} fashion, re-sampling in every iteration is no longer necessary. Nevertheless, we still re-sampled in our numerical experiments (described below) to reduce the sampling bias, and since the classical sampling procedure was not computationally costly in our examples.

NSQST with pre-training resembles coordinate descent optimization, with $\lambda_1$ and $\lambda_2$ being the two coordinate blocks. Optimizing $\lambda_1$ first and fixing it for the optimization of $\lambda_2$ reduces the dimension of the parameter space for the optimizers throughout the training. However, this does not guarantee convergence to a better local minimum in the loss landscape. We do not intend to demonstrate a clear advantage for NSQST with pre-training over the standard NSQST protocol, as the former uses more computational resources both experimentally (by requiring more measurements) and classically (in the form of the memory, time, and energy consumed to train the neural network quantum state). See \cref{Appendix_hyper} for a comparison of the number of model parameters and the number of measurements used for each of the two approaches. Another motivation for introducing this modified NSQST protocol is to provide new perspectives on the differences between learning the probability amplitudes and the phases of a target quantum state, as well as to inspire other useful hybrid protocols in the future.

\section{Numerical simulations without noise} \label{sec:noiseless_sim}

In this section, we first demonstrate the advantage of our NSQST protocols over the NNQST protocol in three physically relevant scenarios, then demonstrate advantages over direct shadow extimation. Specifically, we consider a model from high-energy physics (time evolution for one-dimensional quantum chromodynamics), a model from condensed-matter physics (time evolution of a Heisenberg spin chain), and a model relevant to precision measurements and quantum information science (a phase-shifted GHZ state).

For all three physical settings, we compare the performance of NNQST, NSQST, and NSQST with pre-training by measuring the exact infidelity of the trained model quantum states to the target state averaged over the last 100 iterations of training (or epochs for NNQST, see \cref{Appendix_hyper}). For NNQST's basis selection, since none of our target states is known to be the ground state of a $k$-local Hamiltonian (i.e., a Hamiltonian with each term acting non-trivially on at most $k$ qubits), we simply use all of the almost-diagonal and nearest-neighbour local Pauli bases (i.e., Pauli bases with at most two neighbouring terms being non-$Z$). The number of these bases scales linearly with the system size ($4n-3$ bases). All NSQST protocols use only $N=100$ re-sampled classical shadows per iteration for model parameter updates. We perform ten independent trials of each protocol (NNQST, NSQST, and NSQST with pre-training) in each of the three examples. 

Finally, we adopt an improved pre-training strategy described in \cref{app_technical} and fix $N=200$ Clifford shadows without re-sampling to demonstrate advantages of NSQST over direct shadow estimation with Clifford shadows or Pauli shadows.

\subsection{Time-evolved state in one-dimensional quantum chromodynamics}\label{su3}

Quantum chromodynamics (QCD) studies the fundamental strong interaction responsible for the nuclear force \cite{marciano1978quantum}. Lattice gauge theory, an important non-perturbative tool for studying QCD, discretizes spacetime into a lattice and the continuum results can be obtained through extrapolation \cite{kogut1983lattice}. Although lattice gauge theory has been extremely successful in QCD studies, simulations of many important physical phenomena such as real-time evolution are still out of reach due to the sign problem in current simulation techniques. Quantum computers are envisioned to overcome this barrier in lattice gauge theory-based QCD simulations and they may open the door to new discoveries in QCD \cite{banuls2020simulating, nachman2021quantum,dalmonte2016lattice, martinez2016real}.

\begin{figure}[!htb]
\centering
\includegraphics[width = \columnwidth]{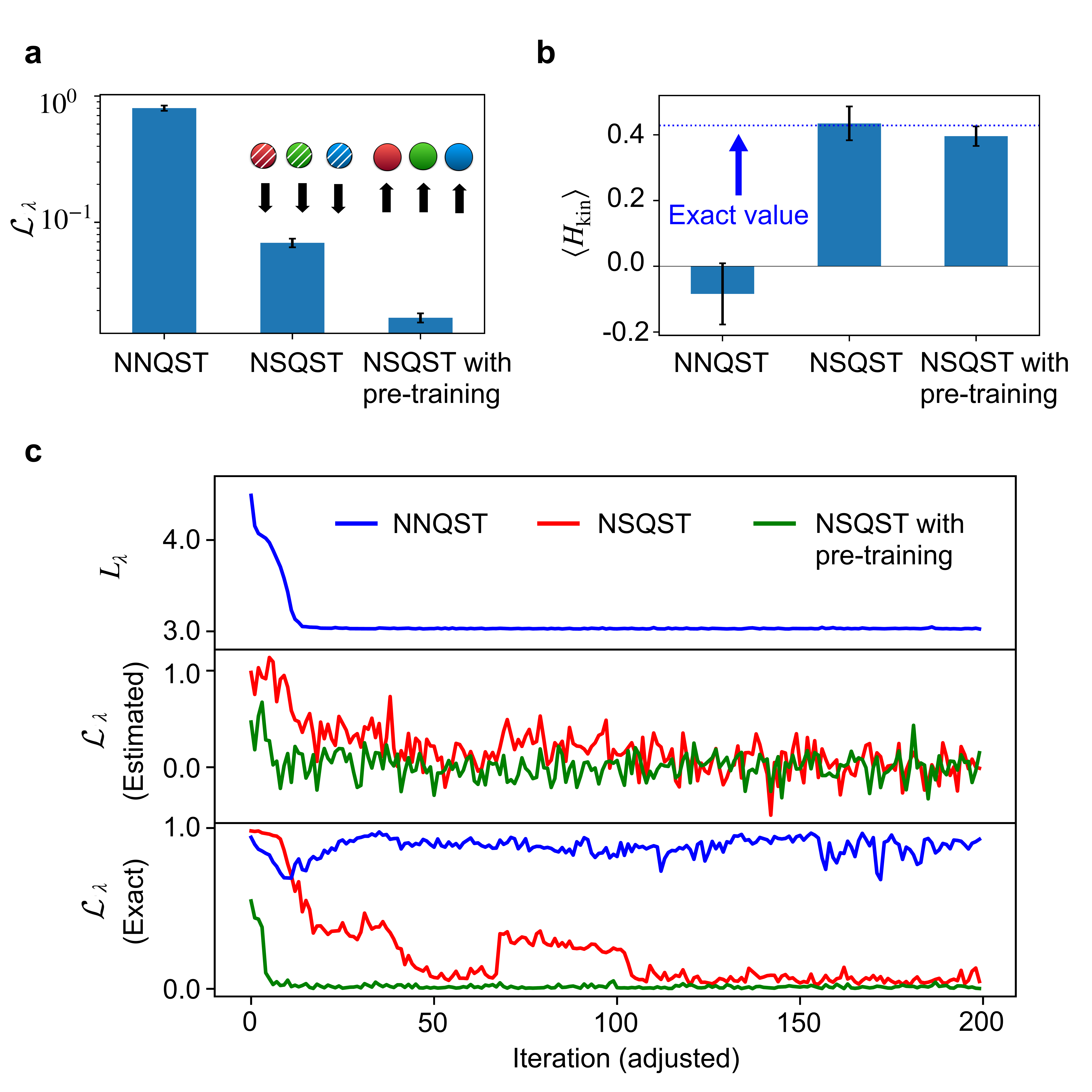}
\caption{Tomography of the quantum state following a one-dimensional QCD time evolution. Two Trotter steps are used for a total evolution time of $t=1.8$. Panel \textbf{a} displays the average final-state infidelity for each of the three protocols. In each trial, we extract the (exactly calculated) average state infidelity $\mathcal L_\lambda$, averaged over the last 100 iterations (and further averaged over ten trials). The error bar is the standard error in the mean calculated over the ten trials. The embedded schematic shows the qubit encoding for a unit cell, containing up to three quarks (filled circles) and up to three antiquarks (striped circles). In panel \textbf{b}, the plot compares the expectation value of the kinetic energy, evaluated for the neural network quantum state found in the last iteration of each trial, and averaged over ten trials for each protocol. In panel \textbf{c}, the optimization progress curves are displayed for a typical trial, where the adjusted iteration refers to epochs for NNQST, but rather indicates increments of ten iterations for the two NSQST protocols (a total of 2000 iterations were run in these cases). Panel \textbf{c} shows the NNQST (blue) loss $L_{ \lambda}$ in the top plot, the estimated NSQST (infidelity) loss function $\mathcal L_\lambda$ (with and without pre-training, green and red, respectively) in the middle plot (fluctuations are dominated by the finite number $N=100$ of classical shadows taken for each estimate), and the exact infidelity is shown in every (adjusted) iteration for all three protocols in the lower plot.}
\label{fig:SU3_1}
\end{figure}

We consider a Trotterized time evolution with the gauge group SU(3) and aim to reconstruct the time-evolved quantum state after a given amount of time. To this end, we use the qubit formulation in Ref.~\cite{atas2023simulating} and study a single unit cell of the lattice. This corresponds to $n=6$ qubits representing three quarks (red, green, blue) and three anti-quarks (anti-red, anti-green, anti-blue) as shown in \cref{fig:SU3_1}a. The Trotterized time evolution starts from the initial state $\ket{\Phi_{0}}=\ket{\downarrow\downarrow\downarrow }\ket{\uparrow\uparrow\uparrow }$, which is known as the strong-coupling baryon-antibaryon state. The Hamiltonian governing the evolution is (from Ref.~\cite{atas2023simulating}):
\begin{equation}\label{su3_ham_1}
{H}_{SU(3)}={H}_{kin}+\tilde{m}{H}_{m}+\frac{1}{2x}{H}_{e},
\end{equation}
where
\begin{equation}\label{su3_ham_2}
\begin{split}
    {H}_{kin}&=-\frac{1}{2} ( {\sigma}_{1}^{+}{\sigma}_{2}^{z}{\sigma}_{3}^{z} {\sigma}_{4}^{-} - {\sigma}_{2}^{+}{\sigma}_{3}^{z}{\sigma}_{4}^{z} {\sigma}_{5}^{-}\\ & \qquad\qquad\qquad\qquad + {\sigma}_{3}^{+}{\sigma}_{4}^{z}{\sigma}_{5}^{z} {\sigma}_{6}^{-} + \operatorname{H.c.}), \\
    {H}_{m}&=\frac{1}{2}\left(6- {\sigma}_{1}^{z}-{\sigma}_{2}^{z}-{\sigma}_{3}^{z}+{\sigma}_{4}^{z}+{\sigma}_{5}^{z}+{\sigma}_{6}^{z} \right), \\
    {H}_{e}&=\frac{1}{3}\left(3-{\sigma}_{1}^{z}{\sigma}_{2}^{z}-{\sigma}_{1}^{z}{\sigma}_{3}^{z} -{\sigma}_{2}^{z}{\sigma}_{3}^{z}\right),
\end{split}
\end{equation}
with $\tilde{m} = am$ and $x = 1/(ga)^2$ for a lattice spacing $a$, where $m$ is the bare quark mass and $g$ is the gauge coupling constant. We use two Trotter steps in our simulation, each for time $t=1.8$. See \cref{Appendix_trotter} and Ref.~\cite{atas2023simulating} for more details on the circuit and the physical significance of this time evolution.

\Cref{fig:SU3_1} shows the results of simulating tomography on the time-evolved state $\ket{\Phi_{SU(3)}}$ using NNQST, NSQST, and NSQST with pre-training.  Note that, although the NSQST protocols (with and without pre-training) are run for 2000 iterations, we use increments of ten iterations in the plot to provide a visual comparison with NNQST (which is run for 200 epochs due to faster convergence of $L_{\lambda}$ in optimization). For NSQST with pre-training, we display the optimization progress curve only after the probabilities $p_{\lambda_1}(s)$ have been pre-trained. This explains the lower initial infidelity for NSQST with pre-training. See \cref{Appendix_hyper} for further details on the simulation hyperparameters.

Based on \cref{fig:SU3_1}, NSQST and NSQST with pre-training both result in a lower final-state infidelity, relative to NNQST, and both predict the mean kinetic energy values better than NNQST. \Cref{fig:SU3_1}c further depicts the optimization progress curves of a typical trial. We see that for NNQST, the cross-entropy loss $L_{ \lambda}$ quickly converges with very little fluctuations, despite the continued fluctuations of the state infidelity in the lower plot near $\mathcal L_\lambda\simeq 1$, indicating a very small overlap with the target state. On the other hand, standard NSQST and NSQST with pre-training both converge to a final state very close to the target state, despite fluctuations in the loss function caused by the finite number of classical shadows in each iteration. Moreover, we notice that NSQST with pre-training not only starts with a state of lower infidelity after pre-training, but also converges to a solution of lower infidelity than standard NSQST, with much more stable convergence in the end. One unexpected outcome is that NSQST with pre-training does not have a better kinetic energy prediction than the standard NSQST despite having lower infidelity. However, this can perhaps be an artifact of insufficient statistics given only ten trials. The predicted total energy and mass are also plotted in \cref{fig:total_eng} (\cref{additional_plots}), where NSQST and NSQST with pre-training yield significantly better predictions of total energy but not the local observable $H_m$.

\begin{figure*}[!htbp]
\centering
\includegraphics[width = 2.0\columnwidth]{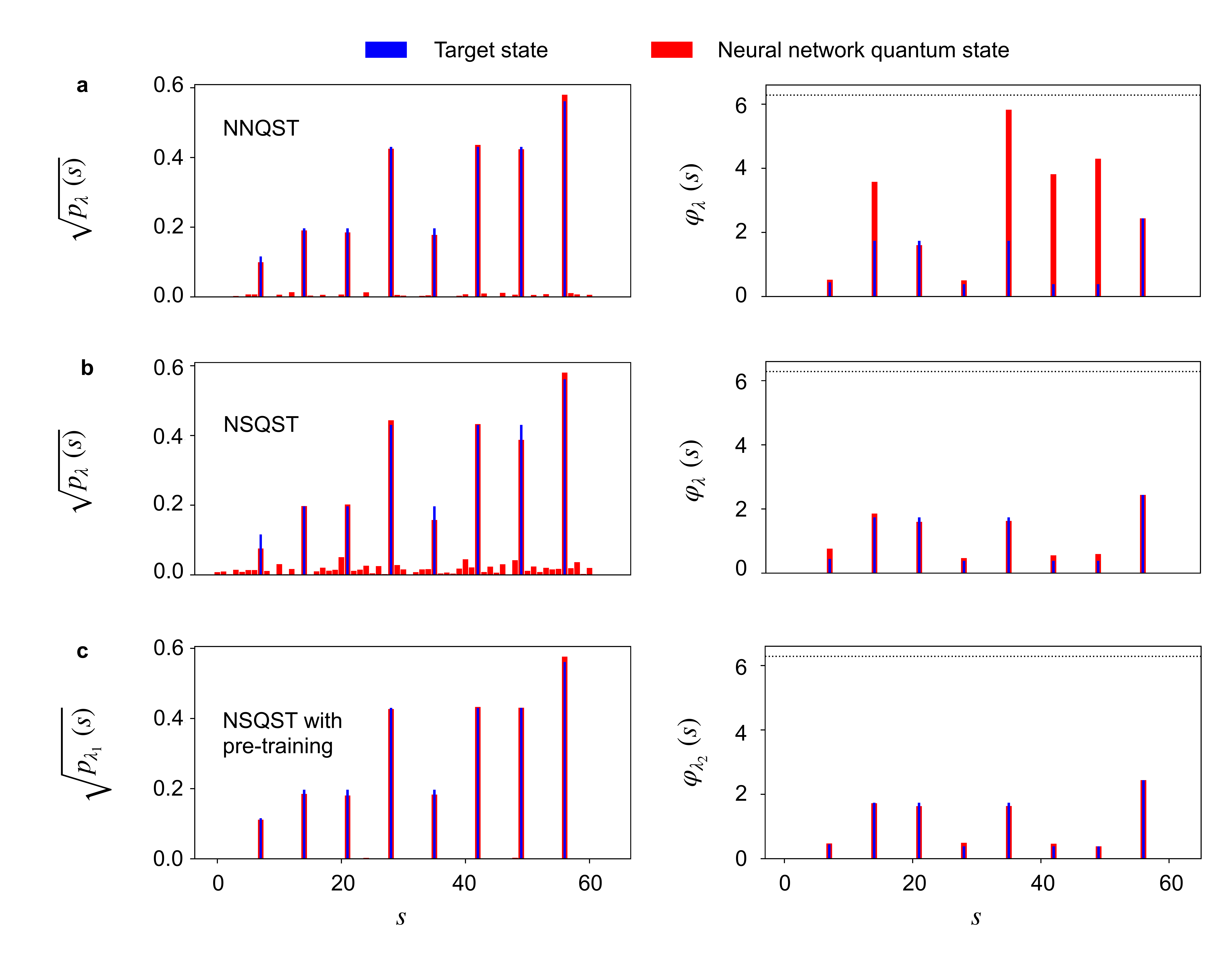}
\caption{Typical neural network quantum states following optimization, approximating the state after a one-dimensional QCD time evolution. Panel \textbf{a} displays a typical state found in the last iteration of NNQST training: the left plot shows the square root of the probability of the final neural network quantum state compared to the exact target state and the right plot shows the phase output of the state over the set of computational basis states $s$. To highlight the dominant contributions, the phase output has been truncated for states $s$ with $\sqrt{p_{\lambda_1}(s)}<0.1$. For the purpose of better visualization, the overall (global) phase of the neural network quantum state is chosen by aligning the phase of the most probable computational-basis state to that of the target state. The dashed line corresponds to a phase of $2 \pi$, since we choose our phase predictions to be in the range $[0, \ 2 \pi]$. Panels \textbf{b} and \textbf{c} show typical final states from NSQST and NSQST with pre-training. We observe that both NSQST protocols succeed at learning the phase structure while NNQST fails at the same task.}
\label{fig:SU3_2}
\end{figure*}

In \cref{fig:SU3_2}, the amplitudes and phases are displayed for typical final neural network quantum states for each protocol. We see that the NNQST protocol fails at learning the phase structure of the target state, despite accurately learning the probability distribution. This observation is consistent with NNQST's convergence to a poor local minimum in the lower plot of \cref{fig:SU3_1}c, with the infidelity values stuck at around $1.0$. On the other hand, standard NSQST and NSQST with pre-training are both successful at learning the phase structure of the target state, while NSQST with pre-training also learns the probability distribution better. 

\subsection{Time-evolved state for a one-dimensional Heisenberg antiferromagnet}\label{AFH}

The Heisenberg model describes magnetic systems quantum mechanically. Understanding the properties of the quantum Heisenberg model is crucial in many fields, including condensed matter physics, material science, and quantum information theory \cite{motta2020determining,gong2014emergent,arovas1988functional}. In this example, we perform tomography on a state that has evolved in time under the action of the one-dimensional antiferromagnetic Heisenberg (AFH) Hamiltonian. We use four Trotter time steps to approximate the time evolution.

The one-dimensional AFH model Hamiltonian is
\begin{equation}\label{AFH_ham_1}
{H}_{AFH}=\sum_{i=1}^{n-1} ({\sigma}^x_i {\sigma}^x_{i+1} + {\sigma}^y_i {\sigma}^y_{i+1} + {\sigma}^z_i {\sigma}^z_{i+1}).
\end{equation}
We choose $n=6$ for our simulation and we take open boundary conditions. The 6-qubit initial state is set to the classical N\'eel-ordered state $\ket{\Phi_{0}}=\ket{\uparrow\downarrow\uparrow \downarrow\uparrow\downarrow }$ and our target state occurs after evolving under the Heisenberg Hamiltonian up to time $t=0.8$. The circuit describing the Trotterized time evolution is given in \cref{Appendix_trotter}.

\begin{figure}[!htb]
\centering
\includegraphics[width = \columnwidth]{Fig_5.pdf}
\caption{Training and results for a simulation of tomography on the time-evolved state for a one-dimensional AFH model. Four Trotter steps are used for a total evolution time of $t=0.8$. Panel \textbf{a} compares the final state infidelity, averaged over ten trials for the three protocols, following the same procedure used for one-dimensional QCD time evolution. Panel \textbf{b} compares the predicted mean staggered magnetization in the x-direction (where ${S}^x_j = \frac{1}{2}{\sigma}^x_j$), following a Trotterized time evolution under the AFH model, for all the three protocols. In Panel \textbf{c}, we show optimization progress curves for a typical run, with the NNQST (blue) loss $L_{ \lambda}$ in the top plot, NSQST loss functions (with and without pre-training, green and red, respectively) in the middle plot, and the exact infidelity in every adjusted iteration for all three protocols in the lower plot.}
\label{fig:AFH_1}
\end{figure}

\Cref{fig:AFH_1} shows the simulation results for performing tomography on the time-evolved AFH state using NNQST, NSQST, and NSQST with pre-training. We see that NSQST and NSQST with pre-training both reach a lower final state infidelity than NNQST in \cref{fig:AFH_1}a. \Cref{fig:AFH_1}b displays the mean staggered magnetization (along $x$) at each site. We observe that NSQST with pre-training results in a tighter spread of values about the exact result, relative to NNQST or NSQST across all sites. Comparing NNQST and standard NSQST, we observe that the standard NSQST protocol has significantly worse predictions at sites 3 and 4 than NNQST, with a mean more than one standard error away from the exact value, despite having a significantly lower final state infidelity. This is likely due to the fact that NNQST is trained using nearly-diagonal measurement data, providing direct access to the staggered magnetization observable of interest, whereas NSQST was trained using the infidelity loss. This result demonstrates that reaching a lower infidelity does not necessarily imply a better prediction of local observables, although we can improve the standard NSQST protocol by using more classical shadows per iteration. Finally, the typical optimization progress curves in \cref{fig:AFH_1}c are consistent with the statistical results shown in \cref{fig:AFH_1}a. For NNQST, the infidelity does not converge stably, despite the convergence of its loss function.

\begin{figure*}[!htb]
\centering
\includegraphics[width = 2.0\columnwidth]{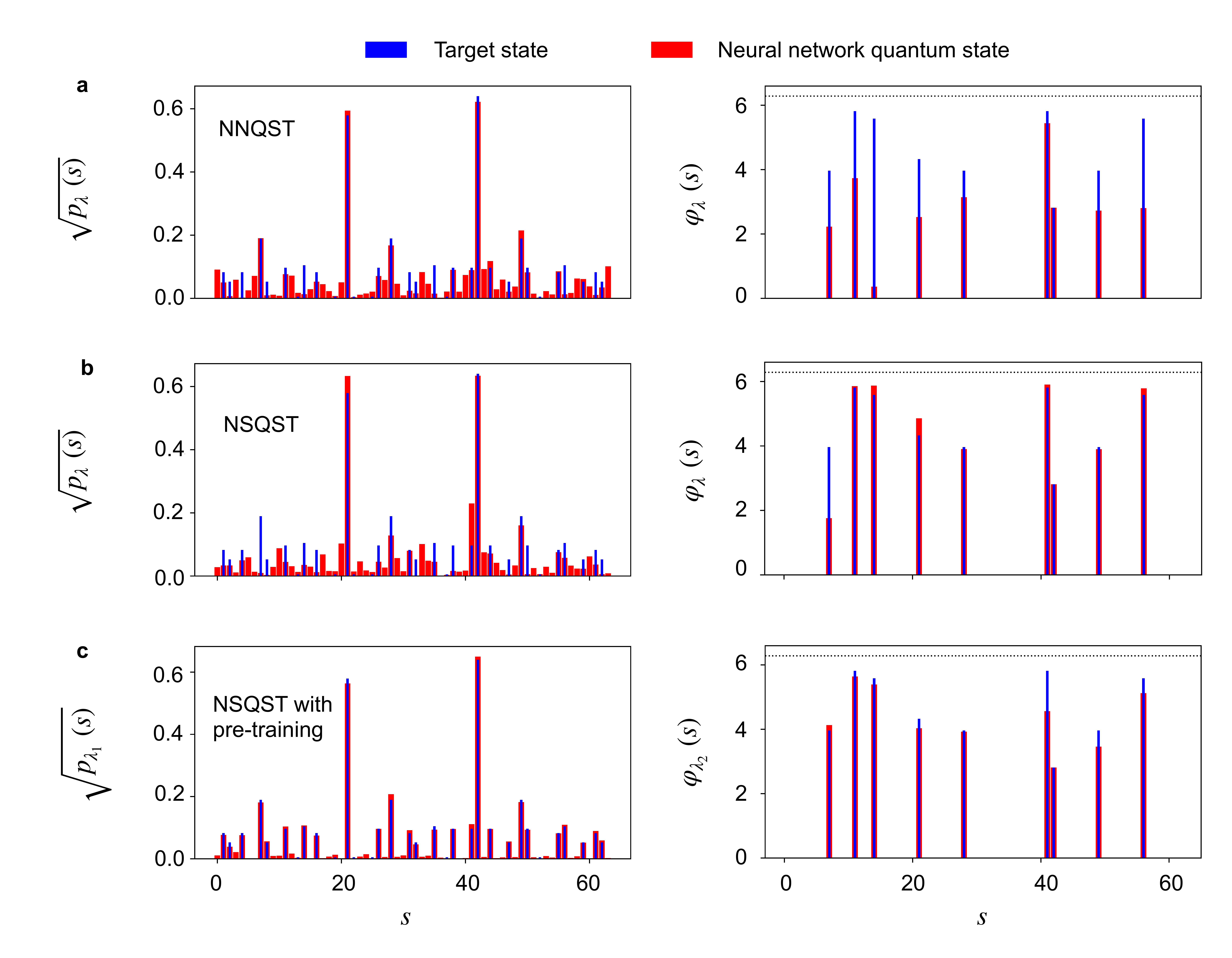}
\caption{Typical final neural network quantum states, trained on the time-evolved state of a one-dimensional AFH model. Panel \textbf{a} displays a typical final state in the last iteration of NNQST optimization, generated using the same procedure from \cref{fig:SU3_2}. Panels \textbf{b} and \textbf{c} show the typical final states from NSQST and from NSQST with pre-training, respectively.  }
\label{fig:AFH_2}
\end{figure*}

The probability amplitudes and phases obtained after training on a time-evolved AFH state are shown in \cref{fig:AFH_2}. Here, the two highest peaks in $p_\lambda(s)$ correspond to the two N$\acute{\text{e}}$el states $\ket{\uparrow\downarrow\uparrow \downarrow\uparrow\downarrow }$ and $\ket{\downarrow\uparrow \downarrow\uparrow\downarrow \uparrow}$ \cite{kennedy1988existence}. \Cref{fig:AFH_2} further confirms the advantage of NSQST with pre-training.  Not only does NSQST with pre-training find more accurate phases, it also finds a better description of the probability distribution since the pre-training involves many measurement samples from the all-$Z$ basis (whereas NNQST splits the same number of measurement samples over multiple bases). 

\subsection{The phase-shifted GHZ state}\label{GHZ}

In this last example, we consider the tomography of a phase-shifted GHZ state. Here, our target is a 6-qubit GHZ state with a relative phase of $\frac{\pi}{2}$. A GHZ state is a maximally entangled state that is highly relevant to quantum information science due to its non-classical correlations \cite{zukowski1998quest}. Moreover, the GHZ state is the only $n$-qubit pure state that cannot be uniquely determined from its associated $(n-1)$-qubit reduced density matrices \cite{walck2008only}, indicating genuine multipartite entanglement.
\begin{figure}[!htb]
\centering
\includegraphics[width = \columnwidth]{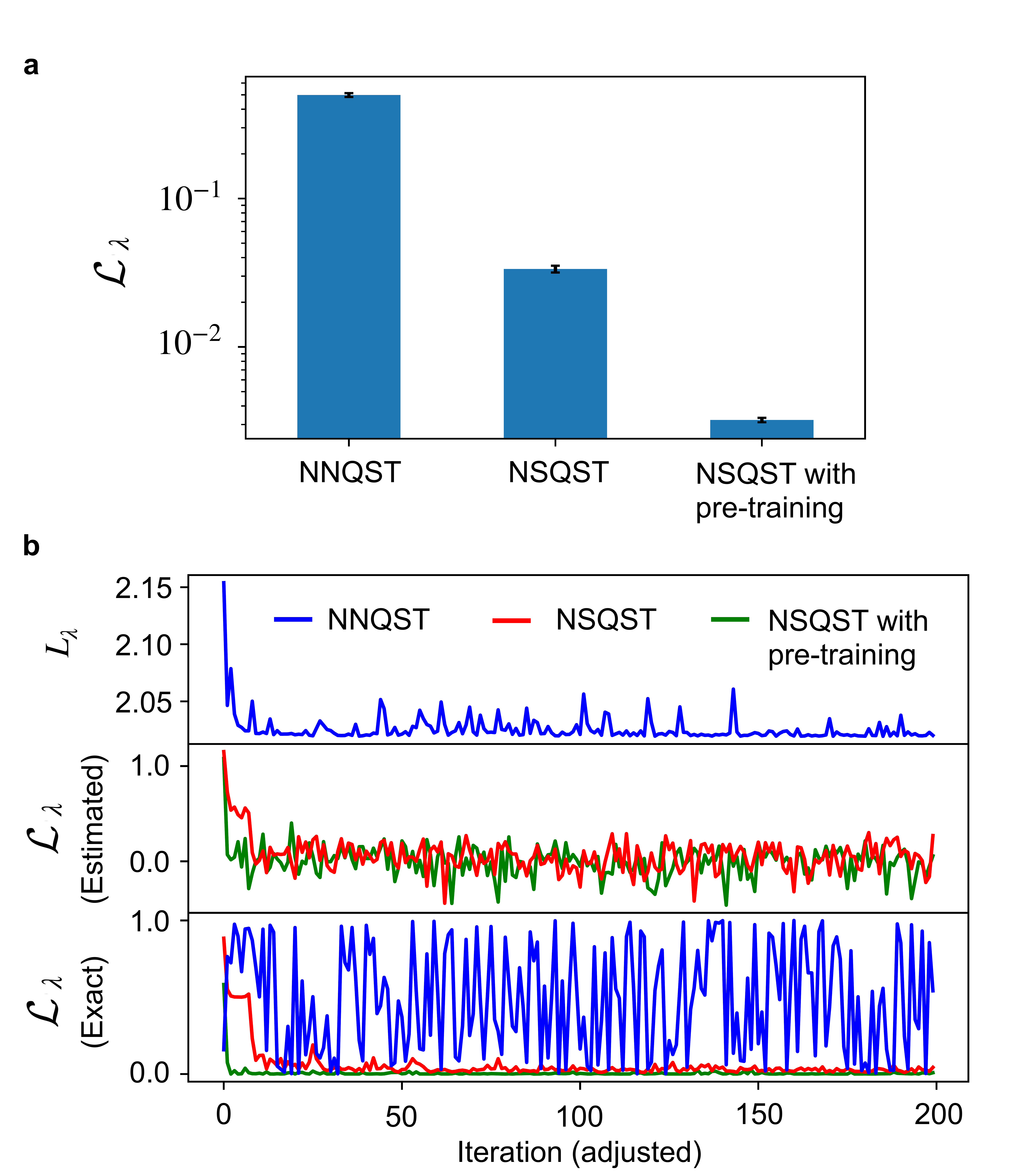}
\caption{Simulation of tomography on a six-qubit phase-shifted GHZ state. Panel \textbf{a} compares the final state infidelity, averaged over ten trials for each of the three protocols. Panel \textbf{b} shows typical optimization progress curves for NNQST (blue), NSQST (red), and NSQST with pre-training (green).}
\label{fig:GHZ_1}
\end{figure}

As shown in \cref{fig:GHZ_1}a, both NSQST and NSQST with pre-training result in a significantly lower average final state infidelity than NNQST. The optimization progress curves displayed in \cref{fig:GHZ_1}b confirm this result, as we see that the infidelity of the NNQST state is rapidly fluctuating during training, quite distinctly from the previous two examples. Given that we have employed a widely used adaptive optimizer and that we have chosen a reasonable initial learning rate ($5 \cdot 10^{-3}$), the occurrence of such a divergence is likely due to the fact that the NNQST loss function does not incorporate tomographically complete information about the target GHZ state.

\subsection{Comparison with direct shadow estimation}\label{direct_shadow_sec}

So far we have only compared the performance of NNQST, NSQST, and NSQST with pre-training, but an important question is whether any of the above methods has an advantage over direct shadow estimation. In this subsection, we compare the performance of NSQST with pre-training and direct shadow estimation for a one-dimensional QCD time-evolved state from \cref{su3}. In addition, we perform a scalability study of the phase-shifted GHZ state with up to $40$ qubits, comparing NSQST to direct shadow estimation. To minimize the number of Clifford shadows used for training, we fix $200$ Clifford shadows as training data and do not re-sample in every iteration. In addition to the original pre-training protocol from \cref{pre_nsqst},  we adopt the improved pre-training strategy described in \cref{app_technical}. A typical optimization progress curve with the improved pre-training strategy is shown in \cref{fig:improved_descent} from \cref{additional_plots}, where very few iterations are required for convergence. Once training is completed, we compare the prediction errors of NSQST with pre-training, Clifford shadows, and Pauli shadows.
\begin{figure}[!htb]
\centering
\includegraphics[width = \columnwidth]{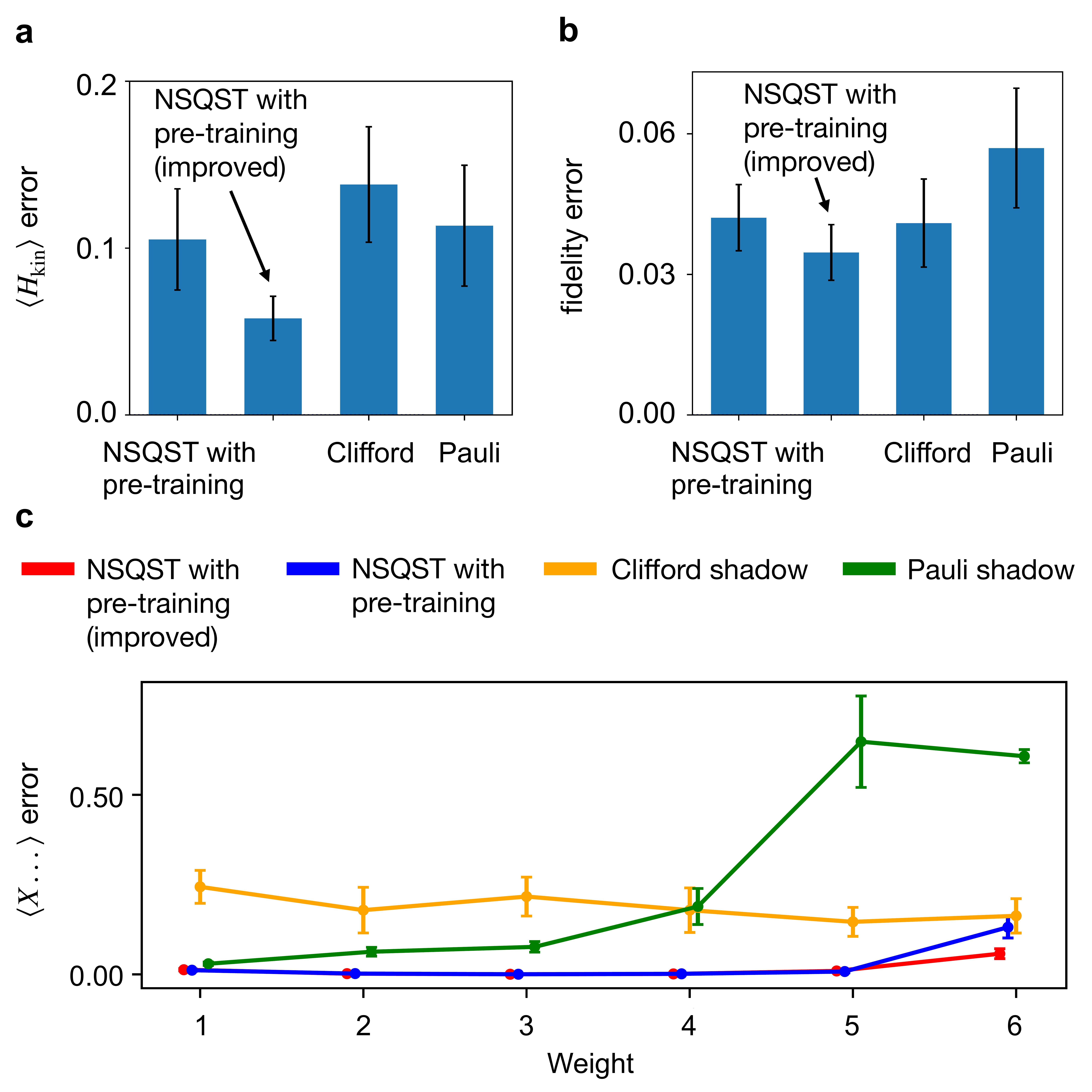}
\caption{Comparison of NSQST with pre-training to direct shadow estimation. In each trial of NSQST with pre-training, $200$ Clifford shadows are used as training data without re-sampling in every iteration and $1000$ measurements in the computational basis are used in pre-training. For direct shadow estimation, $1200$ Clifford shadows and $1200$ Pauli shadows are used. Panel \textbf{a} compares the absolute error in the predicted kinetic energy, averaged over ten trials for each of the four protocols. Panel \textbf{b} compares the absolute error in the predicted fidelity to the ideal time-evolved state, averaged over ten trials for each of the four protocols. Panel \textbf{c} compares the absolute error in the predicted expectation value of a Pauli string observable of various weight in the Pauli-X basis, averaged over ten trials for each of the four protocols. The data points are slightly shifted relative to the ticks of the x-axis for a better display of error bars.}
\label{fig:direct_shadow}
\end{figure}

As shown in \cref{fig:direct_shadow}, we have compared the absolute prediction error of NSQST with pre-training (with and without an improved strategy) and two types of direct shadow estimation methods. For a fair comparison, we have randomly sampled the same number ($1200$) measurements for each method. For shadow reconstruction, $1200$ Clifford shadows or $1200$ Pauli shadows were used. For NSQST with pre-training, $1000$ computational-basis measurements were used for pre-training and $200$ shadows were used. In \cref{fig:direct_shadow}a, we observe that using an improved strategy, NSQST with pre-training achieves a significantly smaller prediction error than either Pauli shadows or Clifford shadows. This is expected, as the kinetic-energy Hamiltonian $H_{\mathrm{kin}}$ in \cref{su3_ham_2} contains high-weight Pauli strings, and Pauli shadows are provably efficient at predicting only local observables \cite{huang2020predicting}. On the other hand, Clifford shadows have an exponentially growing variance bound for any Pauli observables irrespective of locality (since $\tr(O^2) = 2^n$ in \cref{Shadow:scaling_noise}), which explains the large prediction error for kinetic energy. In \cref{fig:direct_shadow}b, we report the predicted error in the fidelity to the ideal time-evolved state. The Pauli shadows yield the largest prediction error in fidelity estimation due to the exponentially growing variance bound for non-local observables. Finally, in \cref{fig:direct_shadow}c, we apply the four methods to the problem of predicting a single Pauli string with increasing weight, where we change the identity matrix to Pauli-X at each site as the weight increases. The non-local observable of interest $\langle X... \rangle$ corresponds to the Wilson loop operator in lattice gauge theory with $\mathbb{Z}_2$ symmetry \cite{zohar2020local}. Since predicting high-weight observables is a hard task for both shadow protocols, the prediction error from NSQST with pre-training is much lower than for either Clifford or Pauli shadows for most of the observables. We also observe a consistent increasing prediction error for the Pauli shadows as the weight increases. 

\begin{figure}[!htb]
\centering
\includegraphics[width = \columnwidth]{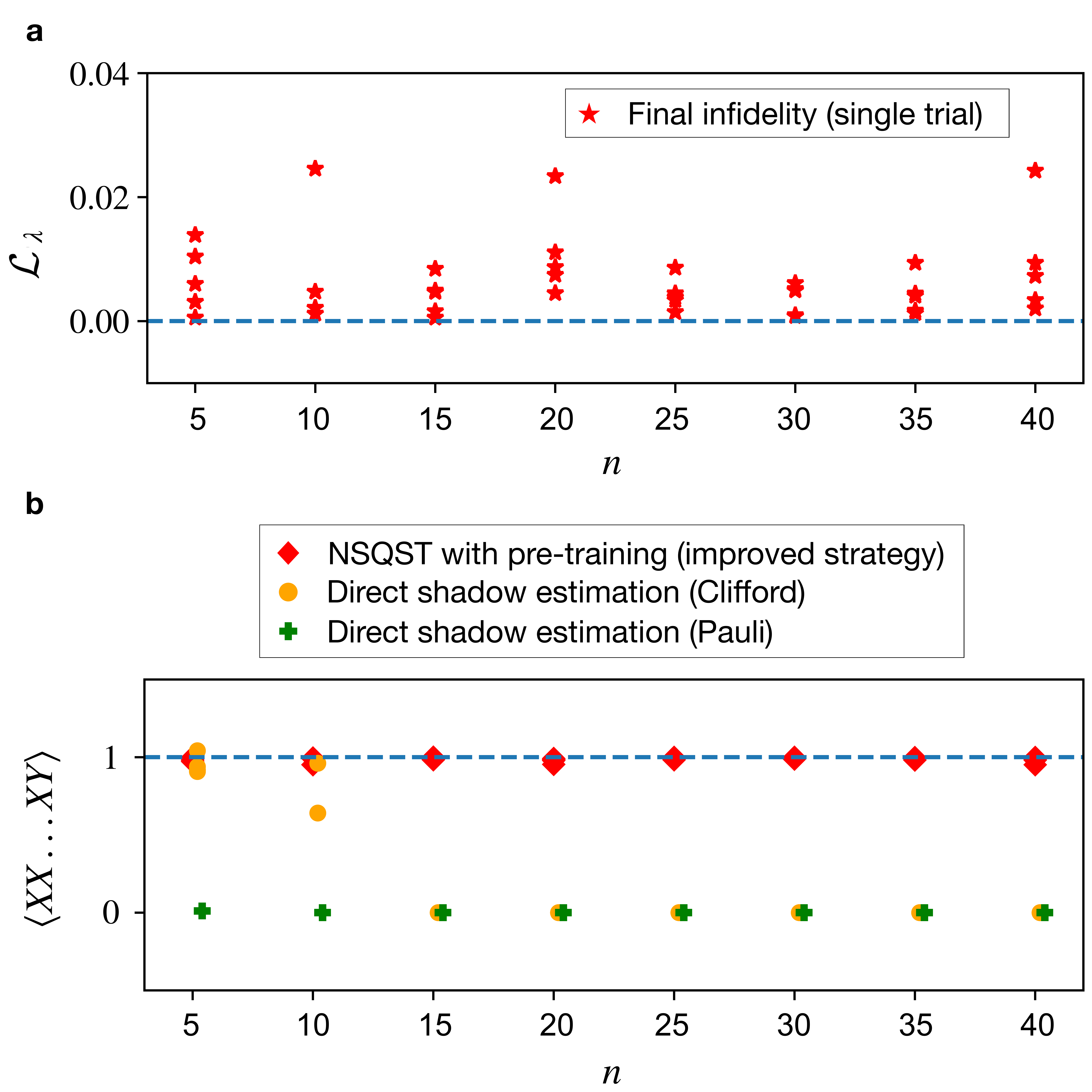}
\caption{Infidelity and predicted expectation value of the phase-shifted GHZ state as the system size grows. For each system size, we generate $3000$ computational-basis measurements and $200$ Clifford shadows. In panel \textbf{a}, with independently sampled measurement data and $5000$ Monte Carlo samples, we run NSQST with pre-training using the improved strategy for five trials and report the individual final infidelities. Note that the number of Monte Carlo samples used during training is much less than the number of basis states, which is not an issue if the target state is sufficiently sparse, as in the case of the multi-qubit GHZ state. In panel \textbf{b}, we plot the predicted expectation value of the Pauli string $ XX...XY $, which is one of the target state's stabilizers. In comparison, the expectation values predicted from five trials of direct shadow estimation are plotted, each with $3200$ independently sampled Clifford or Pauli shadows. The data points are slightly shifted relative to the ticks of the x-axis for a better display.}
\label{fig:scale_up}
\end{figure}
A natural question that arises is the scalability of NSQST's advantages over direct shadow estimation. While a trained neural network quantum state closely approximating the target state has more predictive power than classical shadows alone, there is no guarantee of successful convergence during training. For instance, learning a general multi-qubit probability distribution without any prior knowledge in the pre-training step is hard \cite{blum2003noise} and would eventually require exponentially growing resources. The key to having a scalable advantage is to leverage prior knowledge of the prepared target state and to find ways to impose these known constraints in the neural network ansatz and the loss function \cite{choo2018symmetries, morawetz2021u, haldar2021variational}. 

As a proof of concept, we numerically study the sample complexity scaling of learning a phase-shifted GHZ state using NSQST with pre-training and using the improved strategy from \cref{app_technical}. With $3000$ measurements in the computational basis, $200$ Clifford shadows, and $5000$ Monte Carlo samples for each system size, we investigate the scaling of the final infidelity. As shown in \cref{fig:scale_up}a, the final infidelity does not grow as the system size increases. This is expected, as the multi-qubit GHZ state is sparse in the computational basis and only a single relative phase needs to be determined. Although learning the GHZ state using a neural network ansatz is a trivial example, the sparseness property of the GHZ state is not exploited by direct shadow estimation and the collected Clifford shadows as training data would not be sample-efficient at predicting Pauli observables. In \cref{fig:scale_up}b, we observe that direct shadow estimation fails to predict expectation values $\langle XX...XY \rangle$ accurately and yields only values of zero as the system size increases. The result presented in \cref{fig:scale_up} suggests that there is hope to reconstruct sufficiently sparse states with sub-exponentially growing resources using NSQST, and potentially non-sparse states as well with enough known constraints imposed. Finally, we emphasize that, as compared to the neural network quantum state ansatz trained on IC-POVMs data in Ref.~\cite{carrasquilla2019reconstructing}, our chosen state ansatz explained in \cref{method_NN_state} is sample-efficient at predicting Pauli string observables of arbitrary weight and fidelity to any classically tractable states.

We make a final remark on the predictive power of randomized measurements alone versus a trained variational pure state. While one may generally expect the trained pure state to inherit the features of the training data, this may not be true in specific cases for NSQST and Clifford shadows, where the trained pure state's predictive power mainly depends on the global reconstruction error (quantum infidelity), rather than an estimator for a particular observable. Moreover, the variational training framework of NSQST is not limited to a neural network quantum state with Clifford shadows. Other variational ansatzes such as matrix product states and other randomized measurement schemes such as Hamiltonian-driven shadows should be explored with proper locality adjustments, for practical advantages such as hardware-aware measurements and scalable classical post-processing \cite{hu2022hamiltonian, hu2023classical, cramer2010efficient, lanyon2017efficient}. 

\section{Numerical simulations with noise} \label{sec:noisy_sim}

We now numerically investigate the noise robustness of our NSQST protocol, focusing on the same phase-shifted GHZ state as in \cref{GHZ}. We consider two different sources of noise affecting the Clifford circuit used to evaluate our infidelity-based loss function using classical shadows (see \cref{standard_nsqst}). The first noise model, (a particular case of the model already introduced in \cref{method_shadow}), describes either measurement (readout) errors or gate-independent time-stationary Markovian (GTM) noise. The second noise model describes imperfect two-qubit entangling gates in the Clifford circuit. In the following, we introduce both noise models and discuss their effects on the fidelity of the reconstructed state. 

Our first model is an amplitude damping channel applied before measurements. The amplitude damping channel is suitable for investigating the effect of measurement errors in the computational basis. The $n$-qubit amplitude damping noise channel $\mathrm{AD}_{n,p}$ with channel parameter $p$ is defined as
\begin{equation}\label{n_amp_damp_def_1}
\begin{aligned}
    \mathrm{AD}_{n,p} = \mathrm{AD}_{1, p}^{\otimes n},
\end{aligned}
\end{equation}
where
\begin{align} \label{n_amp_damp_def_2}
\mathrm{AD}_{1,p}: \begin{pmatrix}
\rho_{00} & \rho_{01} \\
\rho_{10} & \rho_{11}
\end{pmatrix}
\mapsto
\begin{pmatrix}
\rho_{00} + (1-p) \rho_{11} & \sqrt{p} \rho_{01} \\
\sqrt{p} \rho_{10} & p \rho_{11}
\end{pmatrix}.
\end{align}
Apart from modeling measurement noise, this noise channel also serves as a suitable model for studying gate-independent, time-stationary, and Markovian (GTM) noise \cite{chen2021robust}. In this case, each gate that appears in the Clifford circuit $U_i$ is subject to the same noise map. The resulting noisy random Clifford circuit $\tilde{U_i}$ can be decomposed into $\mathcal E U_i$ with $\mathcal E$ being a noise channel applied after the ideal Clifford unitary. 

To demonstrate the noise robustness of our NSQST protocol, we first perform tomography on a phase-shifted GHZ state (having a relative phase of $\frac{\pi}{2}$).  Despite the presence of the amplitude damping noise $\mathcal E = \mathrm{AD}_{n,p}$, we simulate NSQST using the noise-free gradient expression $\nabla_\lambda \mathcal L(\mathbb I)$ in \cref{nsqst_grad}. As discussed in \cref{standard_nsqst}, the noise-free gradient expression in NSQST will still yield an estimate that is directed along the true gradient, being modified only with an overall prefactor. In contrast, the noise-free loss function $\mathcal L (\mathbb I)$ and the true loss $\mathcal L(\mathcal E)$ are related nontrivially in the presence of noise:
\begin{equation}\label{loss_transformation}
\mathcal L(\mathcal E)
= \frac{1}{(2^n+1) f(\mathcal E)} \mathcal L(\mathbb I) +
 \frac{(4^n - 1)f(\mathcal E) - 2^n + 1}{2^n (2^n+1) f(\mathcal E)}.
\end{equation}
This means that our estimated loss function no longer converges to zero, while the infidelity between the neural network quantum state and the target state approaches zero during training.

\begin{figure}[!htb]
\centering
\includegraphics[width = \columnwidth]{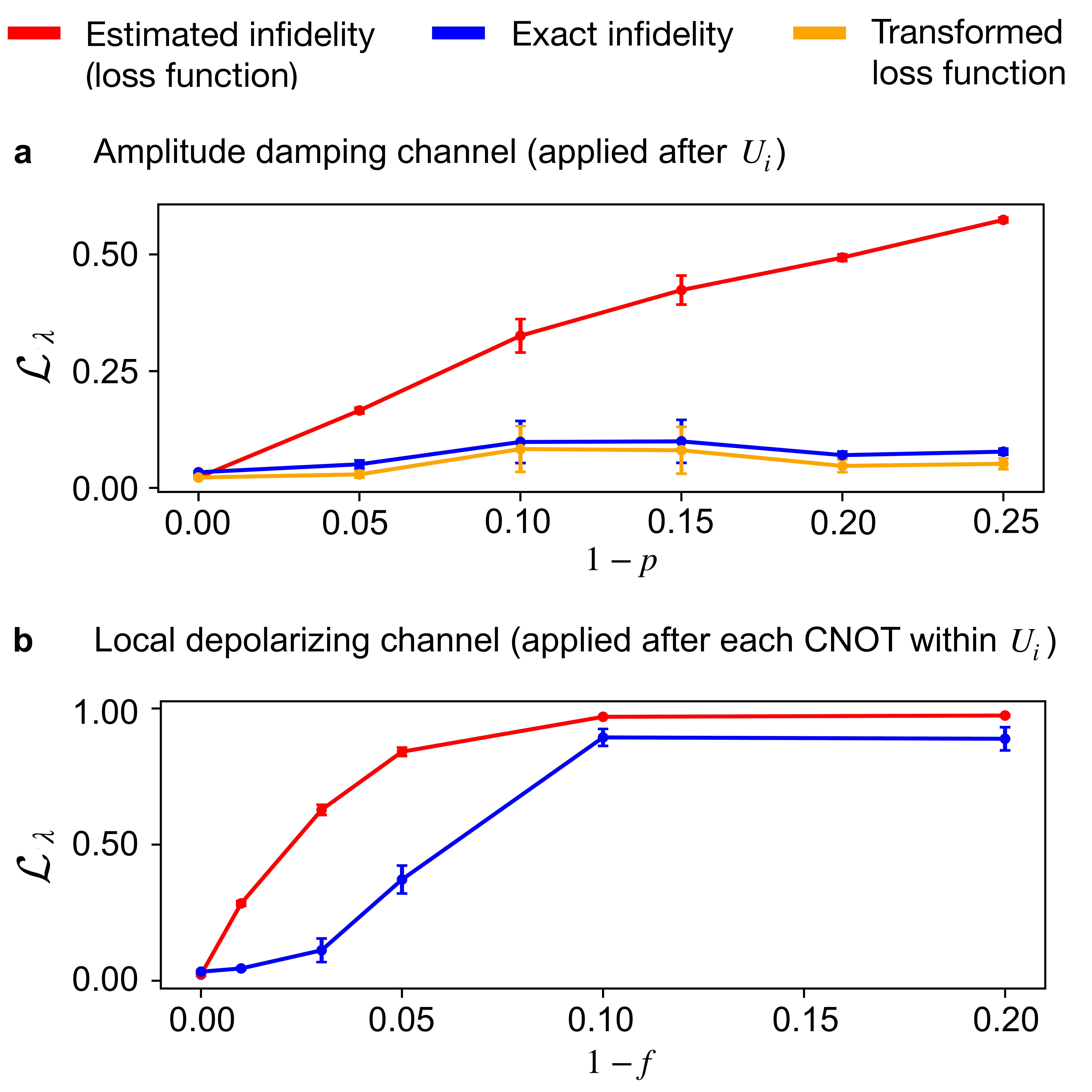}
\caption{Simulation of tomography for a phase-shifted GHZ state in the presence of noise. Panel \textbf{a} displays the average loss function (red) defined in \cref{f_ep_noise_2} and the exact infidelity (blue) for the amplitude damping channel $\mathrm{AD}_{n,p}$ (the loss function is averaged over the last 100 iterations for each trial and then the average is taken over ten trials). The strength of the noise increases with increasing $1-p$. The noiseless infidelity loss function $\mathcal L(\mathbb I)$ is then transformed into an estimated infidelity for the noisy case $\mathcal L (\mathcal E)$ using \cref{loss_transformation} for the amplitude damping channel $\mathcal E=\mathrm{AD}_{n,p}$ to obtain the transformed cost function. The error bars represent the standard error in the mean over ten trials. In panel \textbf{b}, we show the average loss function and exact infidelity for the local depolarizing noise model with a two-qubit depolarizing channel applied after every CNOT gate in the appended random Clifford circuit $U_i$. The channel parameter $1-f$ characterizes the growing strength of the noise. We do not plot the transformed loss function in this case because the CNOT-dependent local depolarizing noise model does not have an analytic noisy shadow expression.}
\label{fig:noise_1}
\end{figure}

\Cref{fig:noise_1}a shows the simulation results for the effects of an amplitude damping noise channel applied before measurement. First, we observe that the average exact infidelity of the last 100 iterations remains small despite the growing noise channel strength. The increasing loss function value (red curves) is evidence of the growing variance in our gradient estimations, and will eventually lead to failure of the optimizer to converge to a state close to the target state. Intuitively, since the classical shadows method only uses the measured bit-string, but not the phase for post-processing, only the diagonal bit-flip errors in \cref{n_amp_damp_def_2} contribute to the noise model and these are twirled into depolarizing noise by random Clifford circuits. Finally, the agreement between the exact infidelity (blue curve) and the transformed loss function (the right hand side of \cref{loss_transformation} represented by the orange curve) validates our theoretical account. Here, we have used \cref{Shadow:f_eps} to find the depolarizing noise channel strength
\begin{align}
    f(\mathrm{AD}_{n,p}) = \frac{(1+p)^n - 1}{2^{2n} - 1} \label{f_ep_noise_2}.
\end{align}
Note that $f(\mathcal E)$ may be hard to estimate in practice. However, since this parameter does not affect the direction of the estimated gradient, we expect training to converge to the same optimal parameters $\lambda$ with or without noise. It is therefore not necessary to compensate for noise by computing the linear transformation in \cref{loss_transformation} as long as we can verify the successful convergence of training.

We proceed now with a discussion of the second noise model, which assumes that entangling gates are the dominant source of error. For numerical simulations, we decompose each random Clifford unitary $U_i$ into CNOT, Hadamard, and phase gate operations. Subsequently, a local two-qubit noise map is applied after each CNOT gate in $U_i$. We consider the depolarizing noise channel
\begin{equation}\label{n_dep_def}
\begin{aligned}
    \mathcal D_{n,f}(\rho) =
    f \rho + (1-f) \frac{\mathbb I}{2^n},
\end{aligned}
\end{equation}
with $n=2$ and a fixed noise strength $1-f$. This noise model is not GTM and no longer has an analytic noisy shadow expression. However, we still expect NSQST to be fairly noise robust based on the numerically demonstrated robustness of classical shadows against many non-GTM errors, such as pulse miscalibration noise \cite{chen2021robust}.

As shown in \cref{fig:noise_1}b, NSQST exhibits some measure of noise robustness even in the presence of a more realistic non-GTM noise model. This is reflected in the positive curvature of the blue curve for decreasing noise, $1-f\to 0$, leading to a weak-noise limit where the exact infidelity (blue curve) is small relative to the estimated infidelity (red curve). Our randomly-sampled six-qubit unitary $U_i$ has an average of 21 CNOT gates (see \cref{Appendix_hyper}), leading to a substantial accumulation of errors. Thus, the noise parameter $1-p$ controlling one-time measurement errors is not comparable to the parameter $1-f$ controlling the noise on the individual CNOT gates. A transformed loss function curve is not presented in \cref{fig:noise_1}b because our local (two-qubit) depolarizing noise model does not yield an analytic $f(\mathcal E)$ expression. The robustness of the classical shadows formalism against many other non-GTM noise models (with an extra calibration step) has been well studied in Ref.~\cite{chen2021robust}, while our NSQST protocol holds a similar noise robustness without any extra calibration steps.

\section{Conclusions and Outlook} \label{sec:conclusions}

In this work, we have proposed a new QST protocol, neural--shadow quantum state tomography (NSQST). We have demonstrated its clear advantages over state-of-the-art implementations of neural network quantum state tomography (NNQST) in three relevant settings, as well as advantages over direct shadow estimation. We have further shown that NSQST is noise robust. Our study of the benefits of NSQST suggests that the choice of infidelity as a loss function has great potential to broaden the applicability of neural network-based tomography methods to a wider range of quantum states.

In \cref{app_technical} we describe technical developments (re-use of classical shadows and alternative Monte Carlo methods) that can be pursued to further enhance the performance of NSQST. Another direction for future work would be to tailor NSQST more closely to emerging quantum hardware platforms. This can be done by exploring NSQST with alternative shadow protocols. In particular, it would be interesting to investigate hardware-aware classical shadows that use the native interactions of the quantum device \cite{van2022hardware, hu2022hamiltonian, hu2023classical}. In addition, future work should extend NSQST to mixed-state protocols \cite{cha2021attention}.

Relative to classical shadow protocols, which only allow for efficient fidelity and local observable predictions, but no efficient state reconstruction, NSQST achieves the goal of reconstructing a physical state that approximates a target quantum state via a variational ansatz. The variational ansatz in NSQST comes with the convenience of a quantum state and can be used to predict many global observables of interest beyond the reach of direct shadow estimation \cite{hibat2023investigating}. Moreover, NSQST inherits the advantages of NNQST. For example, we can incorporate symmetry constraints of the target state, reducing the computational resources needed \cite{luo2021gauge,choo2018symmetries,morawetz2021u}. Once trained, relative to the large number of classical shadows that must be collected, the variational ansatz in NSQST may yield a more efficient classical representation of the state. Finally, as demonstrated in NSQST with pre-training, the trained variational ansatz approximating the target state can be fed into a second round of optimization, performed with respect to a new loss function. This possibility provides great flexibility in addressing a variety of tasks, including, for example, error mitigation in classical post-processing \cite{bennewitz2022neural}.

NSQST is an efficient state reconstruction method. It will be useful as a benchmarking tool, an important element for testing the performance of near-term quantum devices as they scale up. In particular, NSQST can be used to construct a ``digital twin'' \cite{tao2018digital} of the prepared target state, where a neural network quantum state can be used for experimentally-relevant simulation \cite{gutierrez2022real, medvidovic2021classical}, cross-platform verification \cite{elben2020cross}, error mitigation \cite{bennewitz2022neural} and other uses. Having access to a digital twin of the target quantum state will become increasingly relevant for accelerating the development of quantum technologies. Further down the road, we also foresee great potential for NSQST as a stepping stone for interfacing classical probabilistic graphical models and quantum circuits, where data stored in quantum circuits can be transferred to classical memory and vice versa, leading to new hybrid computing approaches.

\medskip

\begin{appendix}

\section{Neural network quantum state architecture}\label{Appendix_nn}

For NNQST and NSQST, we use the transformer-based neural network quantum state ansatz directly adopted from Ref.~\cite{bennewitz2022neural}. A central component in the ansatz is the transformer layer, which has a self-attention block followed by a linear layer. With a bit-string $s = (s_1,\ldots,s_n) \in \{0,1\}^n$ as input, $s$ is extended to $\tilde s = (0, s)$ by prefixing a zero bit. Then, each bit $\tilde s_j$ is encoded into a $D$-dimensional representation space using a learned embedding governed by $f_{jd}$, yielding the encoded bit $e^{(0)}_{jd}$ with $j \in \{0, \ldots, n\}$ and $d \in \{1, \dots, D\}$. The encoded input is then processed using $K$ transformer layers.

We outline the parameters involved in a single transformer layer indexed by $k$ in the ansatz, and refer the reader to Ref.~\cite{bennewitz2022neural} for more details:
\begin{enumerate}
\item The query, key, value matrices $Q^{(k)}_{hid}$, $K^{(k)}_{hid}$, and
$V^{(k)}_{hid}$, where $h \in \{1, \ldots, H\}$, $d \in \{1, \ldots, D\}$, $i \in \{1, \ldots, D/H\}$. We require $D/H$ to be an integer.

\item A matrix to process the output of the self-attention heads,
$O^{(k)}_{de}$, with $d, e \in \{1, \ldots, D\}$.

\item A weight matrix and a bias vector of the linear layer, $W^{(k)}_{de}$ and
$b^{(k)}_d$, with $d, e \in \{1, \ldots, D\}$.
\end{enumerate}

Once we have passed the final transformer layer, scalar-valued logits $\ell_j$ are obtained by using an extra linear layer. The conditional probabilities directly used in sampling are then given by
\begin{equation}\label{cond_prob}
  p_{ \lambda}(s_j = 1| s_1, \ldots, s_{j-1})
    ={} \sigma(\ell_{j-1}),
\end{equation}
where $\sigma(\ell_{j-1})= \tfrac{1}{1+e^{-\ell_{j-1}}}$ is the logistic sigmoid function. Since the outcome at index $j$ is conditional on the preceding indices $\tilde{j} \leq j$, we can efficiently draw unbiased samples from the probability distribution $p_{ \lambda}(s)$ by proceeding one bit at a time. The phase output $\varphi_{ \lambda}(s)$ is obtained by first concatenating the output of the final transformer layer to a vector of length $n$, then projecting the vector to a single scalar value using a linear layer (separate from the linear layer used in obtaining $p_{ \lambda}(s)$).

For NSQST with pre-training, our $p_{ \lambda_1}(s)$ is parameterized in the same way as in standard NSQST, except that we remove the phase output layer. The phase outcome $\varphi_{ \lambda_2}(s)$ is encoded in a separate transformer-based neural network ansatz, where we remove the other linear layer (the one producing scalar-valued logits representing the probability amplitudes). Thus, the encoded quantum state in NSQST with pre-training has its probability distribution and phase output separately parameterized by model parameters $\lambda_1$ and $\lambda_2$, respectively.

\section{Trotterized time evolution circuits}\label{Appendix_trotter}

In this section, we elaborate on the details of Trotterized time evolution circuits for the two noiseless numerical experiments (\cref{su3,AFH}).
\begin{figure*}[t]
\centering
\includegraphics[width = 1.31\columnwidth]{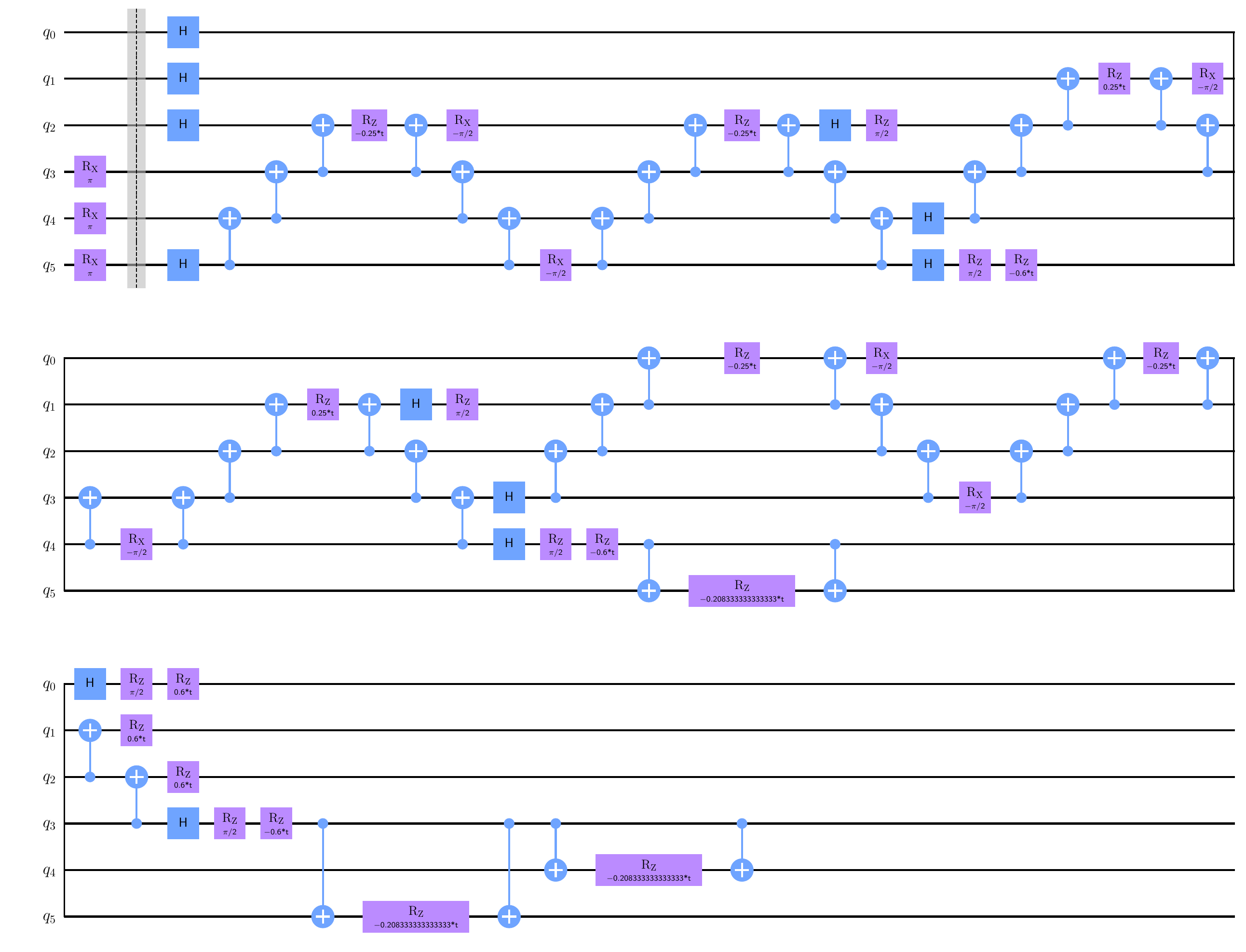}
\caption{Simulation circuit for the one-dimensional QCD model. The initial state preparation circuit (before the barrier) and a single Trotter step (after the barrier) are drawn using Qiskit \cite{aleksandrowicz2019qiskit}. In our numerical experiments an evolution for time $t=1.8$ is decomposed into two Trotter steps.}
\label{fig:su3_cir}
\end{figure*}
\begin{figure*}[t]
\centering
\includegraphics[width = 1.31\columnwidth]{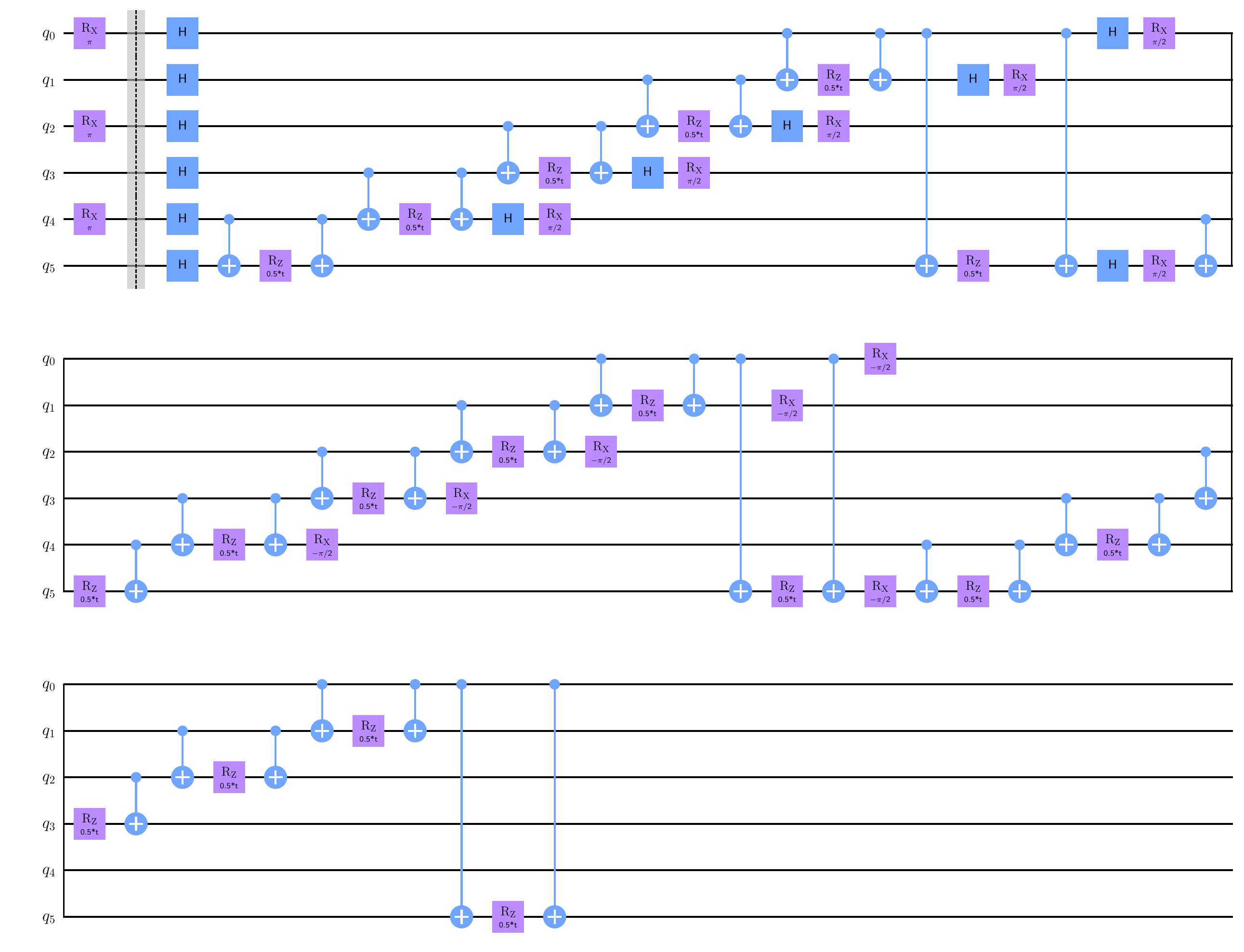}
\caption{Simulation circuit for the one-dimensional AFH model. The initial state preparation circuit (before the barrier) and a single Trotter step (after the barrier) are drawn using Qiskit \cite{aleksandrowicz2019qiskit}. In our numerical experiments an evolution for time $t=0.8$ is decomposed into four Trotter steps.}
\label{fig:AFH_cir}
\end{figure*}

\subsection{One-dimensional QCD time evolution}\label{app_su3}

For the one-dimensional QCD Hamiltonian in \cref{su3_ham_1}, we choose $\tilde{m} = 1.2$ and $x = 0.8$. Two Trotter steps are used for a total evolution time of $t=1.8$. With the initial state chosen as $\ket{\Phi_{0}}=\ket{\downarrow\downarrow\downarrow}\ket{\uparrow\uparrow\uparrow }$, \cref{fig:su3_cir} shows the circuit (from Ref.~\cite{atas2023simulating}) used to generate the time-evolved target state $\ket{\Phi_{SU(3)}}$.

\subsection{One-dimensional AFH time evolution}\label{app_afh}

For the one-dimensional AFH model Hamiltonian in \cref{AFH_ham_1}, we choose the initial state $\ket{\Phi_{0}}=\ket{\uparrow\downarrow\uparrow\downarrow\uparrow\downarrow }$ and set the total evolution time as $t=0.8$, decomposed into four Trotter steps to generate the target state. \cref{fig:AFH_cir} shows the initial state preparation and the circuit of one Trotter step.

\section{Simulation hyperparameters}\label{Appendix_hyper}

For all $n=6$ transformer-based neural network quantum states, we have used two transformer layers ($K=2$), four attention heads ($H=4$), and eight internal dimensions ($D=8$). The neural network quantum states for NNQST and the standard NSQST have $858$ model parameters. NSQST with pre-training uses two neural networks for the same quantum state, where the probability distribution ansatz $p_{\lambda_1}(s)$ has $801$ model parameters and the phase output ansatz $\varphi_{\lambda_2}(s)$ has $849$ model parameters, a total of $1650$ model parameters. It is likely that we have used more model parameters than necessary for a 6-qubit pure state, as the focus of our demonstrations is studying the new loss function. For $n>6$ transformer-based neural network quantum states used for NSQST with pre-training, we have used three transformer layers ($K=3$), four attention heads ($H=4$), and eight internal dimensions ($D=8$), corresponding to a number of model parameters from $2466$ to $3186$ for system size from $n=10$ to $n=40$.

All NNQST protocol trials use 21 ($4n-3$ with $n=6$) nearly-diagonal and nearest-neighbour Pauli bases, where 512 measurement samples are used for each basis. The batch sizes are 128 and all NNQST simulations are run for 200 epochs, where one epoch refers to one sweep over the entire training data set. In all simulations the Adam optimizer is used. For the one-dimensional QCD and AFH time evolution examples, the learning rate is set to $10^{-3}$. For the GHZ state tomography, a learning rate of $5\cdot 10^{-3}$ is used.

All $n=6$ NSQST protocol trials (including the ones with pre-training) use 100 classical shadow samples ($N=100$) and 5000 neural network quantum state samples (both re-sampled in every iteration) per iteration. All simulations are run for 2000 iterations with the learning rate $10^{-2}$ for the Adam optimizer. Note that we have not leveraged the re-usability of the classical shadows, which can bring a significant reduction in the number of measurements. The randomly sampled $6$-qubit Clifford circuits $U_i^{\dagger}$ have $20.8 \pm 0.4$ CNOT gates, assuming all-to-all connectivity.

The $n=6$ modified NSQST experiment with pre-training uses $10752 \ (512\times 21)$ computational basis measurement samples in the pre-training stage using NNQST's optimization framework, with identical hyperparameters to that of the NNQST protocol. NSQST with pre-training (with or without improved strategy) in \cref{direct_shadow_sec} uses $1000$ computational basis measurements for $n=6$ and $3000$ computational basis measurements for $n>6$ in the pre-training stage, and a fixed set of $200$ Clifford classical shadows are used for training the relative phases.

For predicting observable values given a trained neural network quantum state, exact calculations are done for $n=6$ and $20000$ Monte Carlo samples are used for $n>6$. 

\section{Method improvement}\label{app_technical}

For simplicity, we have naively re-sampled the 100 classical shadows in every iteration, which is not necessary as the previously-sampled classical shadows could be re-used. We expect many possible improvements could be made to reduce the number of sampled Clifford circuits and the number of measurements, such as measuring more bit-strings for each $U_{i}$. Hybrid training protocols beyond NSQST with pre-training should also be explored. 
 
 We recognize that an exponential classical computational cost may be induced for convergence due to the growing variance of the gradient in \cref{nsqst_grad}. To intuitively see this, we note that estimating the overlap in \cref{inner_prod} to within $\tilde{\epsilon}$ additive error with failure probability at most $\delta$ requires $L \sim O\left( \frac{1}{\tilde{\epsilon}^2} \log \frac{1}{\delta}\right)$ classical samples drawn from $\lvert \psi_{\lambda} \rangle$ \cite{tang2019quantum}. However, the prefactor $\frac{1}{N f(\mathcal E)}$ in \cref{nsqst_grad} will make the error $\epsilon$ in estimating the gradient exponentially large. Since we want $\epsilon$ to be bounded, we demand $\tilde{\epsilon} \sim N f(\mathcal E)$ to be exponentially small and this directly translates to $L \sim O\left(1/N^2 f(\mathcal E)^2\right)$ as discussed in \cref{standard_nsqst}. Imposing harder constraints on the quantum state ansatz may remove or alleviate this issue, and numerical experiments of larger system sizes should be done to explore NSQST's limitations in the future. 
 
An additional complication can arise when estimating the inner product shown in \cref{inner_prod} from a finite number of Monte Carlo samples. In practice, the overlap is estimated in terms of a subset $\mathcal S$ of distinct bit-strings $s$ using:
 \begin{equation}\label{overlap-appendix}
     \langle\phi_i\ket{\psi_\lambda}=\left<\frac{\phi_i^*(s)}{\psi_\lambda^*(s)}\right>_{\psi_\lambda}\simeq \sum_{s\in\mathcal{S}}P(s)\frac{\phi_i^*(s)}{\psi_\lambda^*(s)}.
 \end{equation}
 Here, $P(s)=f_s/N_{s}\simeq p_\lambda(s)=|\psi_\lambda(s)|^2$ is determined from the frequency $f_s$ of the bit-string $s$ found from $N_{s}$ samples drawn according to the probability distribution $p_\lambda(s)$. For a transformer-based neural network architechture, the samples can be generated efficiently bit-by-bit using the procedure described in \cref{Appendix_nn}. Up to a constant factor, the right-hand side of \cref{overlap-appendix} can be interpreted as the exact overlap between a stabilizer state $\ket{\phi_i}$ and a fictitious state $\ket{\Psi_\lambda}$ with wavefunction:
\begin{equation}\label{unnormalized}
\begin{aligned}
\Psi_{\lambda}(s) := \frac{P(s)}{\psi_{\lambda}^{*}(s)A_{\mathcal S}};\quad A^2_\mathcal{S}=\sum_{s\in\mathcal{S}}\frac{P^2(s)}{|\psi_{\lambda}(s)|^2}.
\end{aligned}
\end{equation}
The normalization constant approaches $A_\mathcal{S}=1$ when $P(s)=p_\lambda(s)=|\psi_\lambda(s)|^2$ (e.g., when the sample set $\mathcal{S}$ includes all $s$). However, a problem arises when we sample only over a subset of possible bit-strings $s$. In this case, it may be that $\lvert \psi_{\lambda}^*(s) \rvert \ll \sqrt{P(s)}$ for some $s$, leading to $A_\mathcal{S}\gg 1$. Estimating the infidelity from classical shadows to obtain the NSQST loss function (\cref{nsqst_loss_noisy}) through Monte Carlo samples as in \cref{overlap-appendix} requires the estimated overlaps $\braket{\phi_i \lvert {\psi_\lambda}}\simeq A_\mathcal{S}\braket{\phi_i \lvert {\Psi_\lambda}}$. When an incomplete sample set is taken, the factor $A_\mathcal{S}$ can become very large, leading to an unphysical blow-up, potentially leading to estimated overlaps $\gg 1$. In this limit, the Monte Carlo estimate is meaningless. A simple solution to this problem could be to truncate the set $\mathcal{S}\to\mathcal{S}'$, allowing only for bit-strings $s$ for which $|\psi_\lambda(s)|/\sqrt{P(s)}$ exceeds some threshold value, then we replace the normalization constant $A_\mathcal{S}\to A_\mathcal{S'}$. For a given task, it may be difficult to establish truncation thresholds that maintain convergence to an accurate state. In the rest of this appendix, we give an alterative procedure that does not show the ill-conditioned ``blow-up'' from a finite Monte Carlo sample size, while avoiding predetermined truncation thresholds.

To avoid the pitfalls of representing a Monte Carlo average as in \cref{overlap-appendix}, we consider a hybrid NSQST protocol. In this hybrid protocol, the classical shadows are only used to learn the phases $\varphi_{\lambda_2}(s)$. The probability amplitudes $p_{\lambda_1}(s)$ are learned using NNQST from measurements performed in the computational basis (similar to NSQST with pre-training): $p_{ \lambda_1}(s)\simeq p_\Phi(s,B)$ with $B = (Z_1, Z_2,\cdots, Z_n)$. The difference between this new hybrid protocol and NSQST with pre-training is in learning the phases. To train the phase model, we calculate the gradient of the loss function with an alternative approximation for the inner product:
\begin{equation}\label{inner_prod_new}
\begin{aligned}
&\langle \phi_{i}\ket{ \psi_{ \lambda}} \approx \langle\phi_i\ket{\widetilde{\psi}_\lambda}:=\sum_{s\in\mathcal S}\phi_{i}^{*}(s)e^{i\varphi_{\lambda_2}(s)} \sqrt{P(s)},
\end{aligned}
\end{equation}
where we have introduced an alternative normalized state:
\begin{equation}\label{normalized}
\begin{aligned}
\widetilde{\psi}_{\lambda}(s) := e^{i\varphi_{\lambda_2}(s)} \sqrt{P(s)}.
\end{aligned}
\end{equation}
The estimated loss function in this new hybrid protocol is the shadow-estimated infidelity between the target state $\ket{\Phi}$ and the state $\ket{\widetilde{\psi}_{\lambda}}$. The gradient of this infidelity can be calculated to optimize the phases from variations in the parameters $\lambda_2$:
\begin{equation}\label{nsqst_grad_new}
\begin{aligned}
\nabla_{\lambda_2} \mathcal L(\mathcal E) \approx \frac{-2}{N f(\mathcal E)} \sum_{i=1}^{N} \Re\Big[ \partial_{\lambda_2} \langle \phi_{i}\ket{ \widetilde{\psi}_{ \lambda}} \cdot \langle \widetilde{\psi}_{ \lambda} \ket{ \phi_i} \Big].
\end{aligned}
\end{equation}
We can efficiently evaluate the right-hand side of \cref{nsqst_grad_new} exactly for a sub-exponential number of distinct bit-strings $s\in\mathcal{S}$. The optimization procedure is then limited only by the expressivity and accuracy of the sparse approximation $\widetilde{\psi}_\lambda(s)$ for the neural network quantum state $\psi_\lambda(s)$, arising from a finite number of samples. This new hybrid NSQST protocol does not suffer from the ``blow-up'' described above and it may converge with a sub-exponential number of samples, especially when $\ket{\psi_{\lambda}}$ is sufficiently sparse in the computational basis. This alternative strategy was unnecessary in most of our numerical experiments given the very small system size $n=6$ and the very large number of Monte Carlo samples $N_s = 5000$. 

\section{Additional plots}\label{additional_plots}

In this section, we provide additional plots relevant to the numerical simulation results presented in \cref{sec:noiseless_sim}. 

\begin{figure}[!htb]
\centering
\includegraphics[width = \columnwidth]{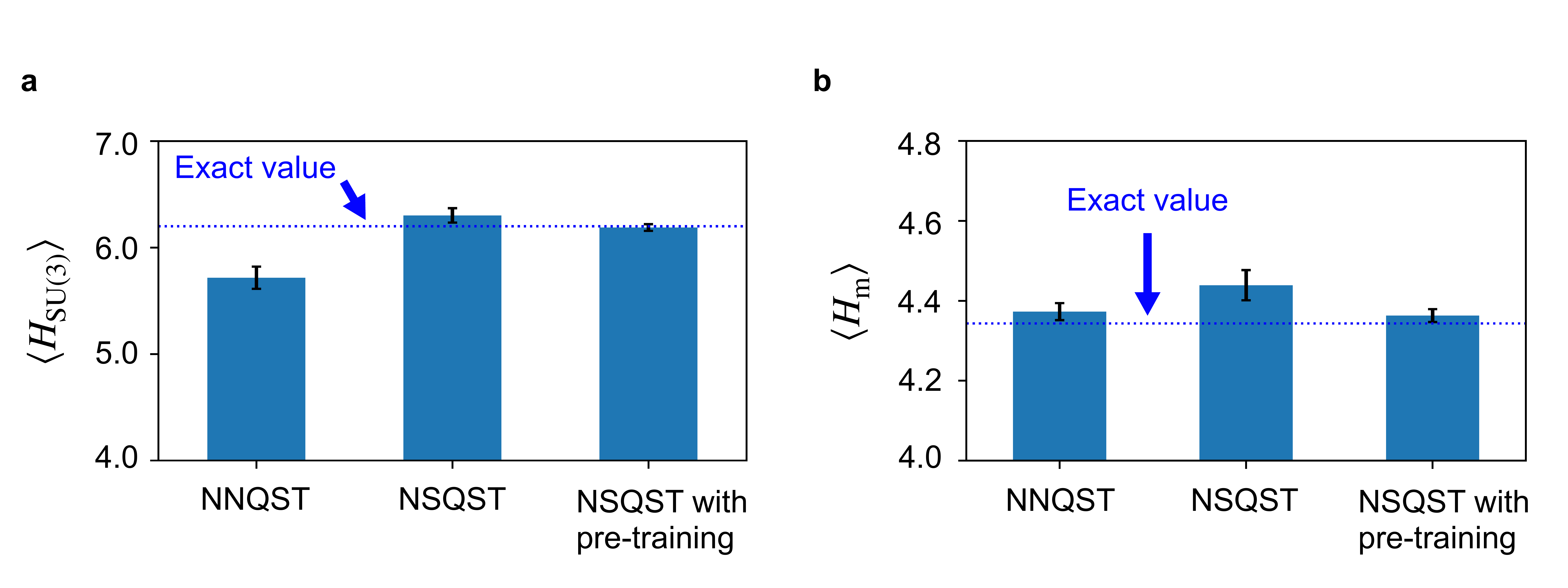}
\caption{Additional plots for the quantum state following a one-dimensional QCD time evolution. Panel \textbf{a} displays the expectation values of the total energy, evaluated for the neural network quantum state found in the last iteration of each trial, and averaged over ten trials for each protocol. Panel \textbf{b} displays the expectation values of the mass Hamiltonian.}
\label{fig:total_eng}
\end{figure}

In \cref{fig:total_eng}, the predicted total energy and mass from the three protocols are plotted, where we see that NNQST fails to yield a better prediction of total energy than NSQST in \cref{fig:total_eng}a. However, as shown in \cref{fig:total_eng}b, NNQST predicts mass $H_m$ more accurately than NSQST, which is a local observable from \cref{su3_ham_2}. 

\begin{figure}[!htb]
\centering
\includegraphics[width = \columnwidth]{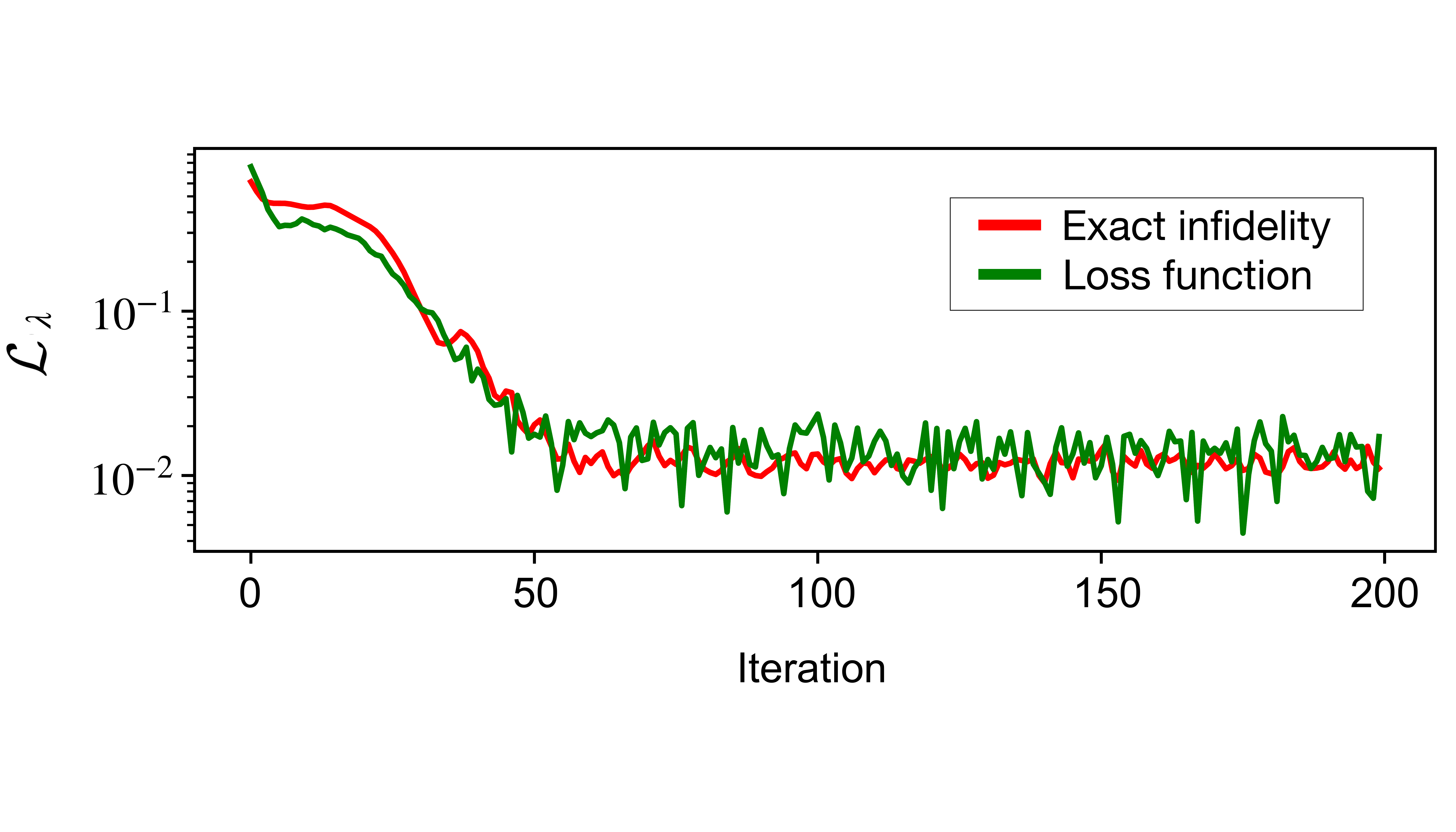}
\caption{Typical optimization progress curve from NSQST with pre-training and fixed Clifford shadows. Unlike the other plots, the iteration number on the x-axis is not adjusted and corresponds to every gradient update during optimization. }
\label{fig:improved_descent}
\end{figure}

In \cref{fig:improved_descent}, a typical optimization progress curve is plotted for the numerical results presented in \cref{direct_shadow_sec}. The iteration number is not adjusted and pre-training is repeated for every trial. 

\end{appendix}

\begin{acknowledgments}
We thank Abhijit Chakraborty, Luca Dellantonio, David Gosset, Arsalan Motamedi, Jinglei Zhang, Jan Friedrich Haase, Yasar Atas, and Randy Lewis for useful discussions. This work has been supported by the Transformative Quantum Technologies Program (CFREF), the Natural Sciences and Engineering Research Council (NSERC), the New Frontiers in Research Fund (NFRF), and the Fonds de Recherche-Nature et Technologies (FRQNT). CM acknowledges the Alfred P. Sloan foundation for a Sloan Research Fellowship and an Ontario Early Researcher Award. PR further acknowledges the support of NSERC Discovery grant RGPIN-2022-03339, Mike and Ophelia Lazaridis, Innovation, Science and Economic Development Canada (ISED), 1QBit, and the Perimeter Institute for Theoretical Physics. Research at the Perimeter Institute is supported in part by the Government of Canada through ISED, and by the Province of Ontario through the Ministry of Colleges and Universities.

\end{acknowledgments}

\medskip 
\medskip

\section*{Code Availability Statement}
The numerical implementation of NNQST from Ref.~\cite{bennewitz2022neural} can be found at \url{https://github.com/1QB-Information-Technologies/NEM}. The numerical implementation of NSQST can be found at \url{https://github.com/victor11235/Neural-Shadow-QST}.

\clearpage
\bibliography{biblio}


\end{document}